\documentstyle[epsf,11pt]{article}

\newcommand{\be}{\begin{equation}}
\newcommand{\ee}{\end{equation}}

\begin{document}
\title{Some properties of deformed Sine Gordon models}

\author{
Akmaral Alibek\thanks{e-mail address: aamaral@mail.ru},\,
Ratbay Myrzakulov\thanks{e-mail address: cnlpmyra1954@yahoo.com},\,
\\
Euroasian University, Astana, Kazakhstan\\
W.J. Zakrzewski\thanks{e-mail address: W.J.Zakrzewski@durham.ac.uk}
\\
Department of Mathematical Sciences, University of Durham, \\
Durham DH1 3LE, UK\\
}
\date{}
\maketitle

\begin{abstract}

We study some properties of the deformed Sine Gordon models. These models, presented by Bazeia et al,
are natural generalisations of the Sine Gordon models in (1+1) dimensions. There are two classes of them,
each dependent on a parameter $n$. For special values of this parameter the models reduce to the Sine Gordon 
one; for other values of $n$ they
can be considered as generalisations of this model. The models are topological and possess one kink solutions.
Here we investigate the existence of other solutions of these models - such as breathers. The work is numerical and we find that
the breathers, as such, probably do not exist. However, we show that some of these models, namely, 
the $n=1$  of the first class
possess breather-like solutions which are quasi-stable; {\it ie} these `quasi-breathers' exist for long periods of time
(thousands of periods of oscillations). These results are found to be independent of the discretisation used
in the numerical part of our work.

\end{abstract}

PACS numbers: 05.45.Ac, 05.60.Cd

\section{Introduction}
Recently \cite{Bazeia} Bazeia et al have presented two classes of models in (1+1) dimensions which are
 natural generalisations of the well known Sine Gordon model. The models in each class depend on one 
parameter $n$. For special values of this parameter, namely $n=2$ for the first class
and $n=1$ for the second class, the models reduce to the Sine Gordon one; for other values of $n$ they
can be considered as generalisations of this model.
The models are defined by the Lagrangian 
\begin{equation}
L=\frac{1}{2}\partial_{\mu}\varphi\partial^{\mu}\varphi\,-\,V(\varphi),
\label{lag}
\end{equation}
where for the first class
\begin{equation}
V\,=\,\frac{2}{n^{2}}\tan^{2}\left(\varphi\right)\left(1-\,sin^{\,n}\left(\varphi\right)\right)^{2}
\label{potone}
\end{equation}
while for the second class
\begin{equation}
V\,=\,\frac{1}{2\,n^{2}}\varphi^{2-2\,n}\,sin^{2}\left(\varphi^{n}\right)
\end{equation}

When $n=2$ for the first class, and $n=1$ for the second, the models reduce to the Sine Gordon model
(for $\varphi=\phi$ or $2\phi$ respectively).

 The models are topological and for any $n$ they posses one kink solutions.

For the first class of models these one kink solutions are given by
\begin{equation}
\varphi\left(\,x,\,t\right)=\,sin^{-1}\left[\frac{\,exp\left(2 \gamma\left(\,x-x_{0}-\,u\,t\right)\right)}{1+\exp\left(2  \gamma\left(\,x-x_{0}-\,u\,t\right)\right)}\right]^\frac{1}{\,n}
\end{equation}

These kink solutions satisfy  the usual Sine- Gordon boundary conditions, namely: 
\begin{equation}
\varphi\left(\,x\rightarrow-\infty\right)\longrightarrow0, \quad \varphi\left(\,x\rightarrow\infty\right)\longrightarrow\frac{\pi}{2}
\end{equation}

The kink solutions of the second class are given by
\begin{equation}
 \varphi\left(\,x,\,t\right)=\left[2\,tan^{-1}\left(\,exp\left( \gamma\left(\,x-\,x_{0}-\,u\,t\right)\right)\right)\right]^\frac{1}{\,n}.
 \end{equation}
Their boundary conditions are 
\begin{equation}
 \varphi\left(\,x\rightarrow-\infty\right)\longrightarrow0, \quad \varphi\left(\,x\rightarrow\infty\right)\longrightarrow\pi^\frac{1}{n}
 \end{equation}
In these expressions $\gamma$ is the usual relativistic factor $\frac{1}{\sqrt{1-u^2}}$, where $u$ 
is the velocity of the kink.

The kink solutions have been studied in the original paper of Bazeia et al \cite{Bazeia} and in a recent paper
by al Alawi et al \cite{Jassem}.
These papers showed that as these models are topological the kink solutions are stable. When the kink solutions
were sent towards various obstructions, in the form of a barrier or a potential hole, the behaviour
of the kinks resembled the behaviour of the kinks of the Sine Gordon model; {\it ie} the kinks could get trapped
in the holes and at some very precise velocities could even get reflected by the holes.

When one considered the dependence on $n$ of these properties (like the critical velocity below which the kinks
were trapped) it was clear that the Sine Gordon model was not special from this point of view; for example,
the critical velocity for the transmission was lowest for $n=3$, and not $n=2$,  in the first class of models.

Hence it is natural to check whether the models of Bazeia et al share other properties of the Sine Gordon model, like integrability
and/or the existence of other fine energy solutions. Of these finite energy solutions the easiest to study
are the multi-kink solutions and breathers.

Unfortunately, so far we have failed to find such solutions analytically. So we have been forced to look 
at them numerically.

In this paper we report some results of our numerical studies. 
In the next sections we present some results of our studies of breathers.
These results were obtained in the usual discretisation.
Our simulations were performed using 4th order Runge-Kutta method of simulating
time evolution. The spatial lattice streched from $-50$ to $50$ and the lattice spacing
was $dx=0.01$. The time evolution spacing $dt$ was taken as $dt= 0.0001$.

Later we look at the topological discretisation 
of Speight and Ward \cite{Ward} and apply it to our case.
We find that that the results are not very different from what we have seen 
using the usual discretisation so that
the fact that our `breather-like' states are long-lived but not
really stable is not a numerical artefact.  In the following section we say a few words
about two kink states.
The final section contains some further comments and a discussion of our future plans.

\section{Breathers}

In the Sine Gordon model ({\it ie} for $n=2$) 
the potential is given by
\be
V(\varphi)\,=\, \frac{1}{8}\sin^2(2\varphi)
\ee
and one has analytical expressions for breathers. For $n\ne2$ the equations of motions 
are more complicated and, so far, we have not found analytically any breathers.
However, as is well known \cite{Johnson}, breathers are bound states of kinks and antikinks and, as such, can
be also found by taking a configuration consisting of a kink and antikink (not too far and not too close 
to each other) and letting it evolve with time. The field quickly readjusts itself to that of a breather
with some small deformations which move out towards the boundaries.
These deformations represent an extra energy which is being sent towards the boundaries.
We absorb it and the field becomes that of the breather.

In fig. 1 we present the energy of three such evolutions for the case of a kink/antikink configuration 
in the Sine Gordon model. We generate such a configuration by taking our field $\varphi$ corresponding 
to a kink located at $x=-d$ for $x<0$ and the corresponding antikink (located at $x=d$) for 
$x>0$. For $d>4.5$ the resultant configuration is quite smooth around $x=0$, but to make it even smoother
we have taken the average values of kink and antikink for $-0.5<x<0.5$.
We have checked that the results do not depend on this `smoothing' procedure.
Here we present the results of three such simulations. In fig 1a - the present the total energy for 
the case of $d=3$, in fig 1b - $d=5$ and in fig 1c of $d=8$. We note only a small initial decrease  in energy
after which the energy is well conserved. This initial decrease is associated with the rearrangement
of the field (from our initial configuration to that of the breather one) and then the field evolves as a 
genuine breather. In fig 2 we present the plot of the potential energy (which for a breather 
varies with time) and in fig 3 we plot the time dependence of the position of the local maximum of the 
energy density in the region of $x\le 0$ (which for a breather corresponds, alternatively to a kink and 
antikink). Of course, we could have given similar plot of the complementary kinks/antikinks
which exist for $x\ge 0$. Our plots demonstrate that our initial configuration 
has turned into a breather, and as one can easily check - the frequency of the breather
(which is related to its `extent') is determined by the initial configuration ({\it ie} its $d$).

\begin{figure}[htbp]
\unitlength1cm \hfil
\begin{picture}(14,8)
 \epsfxsize=7cm \put(-3,0){\epsffile{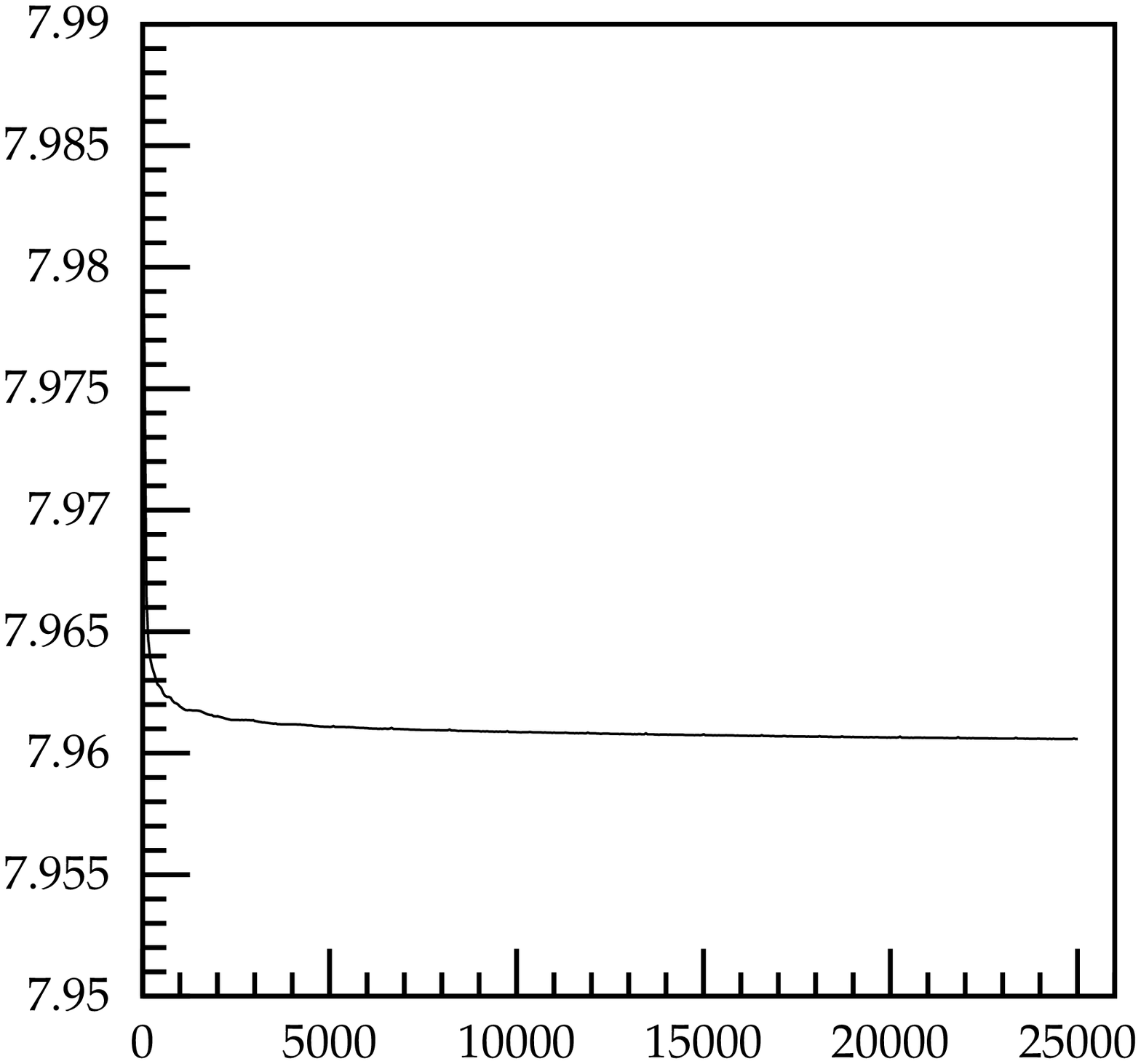}}
 \epsfxsize=7cm \put(3,0){\epsffile{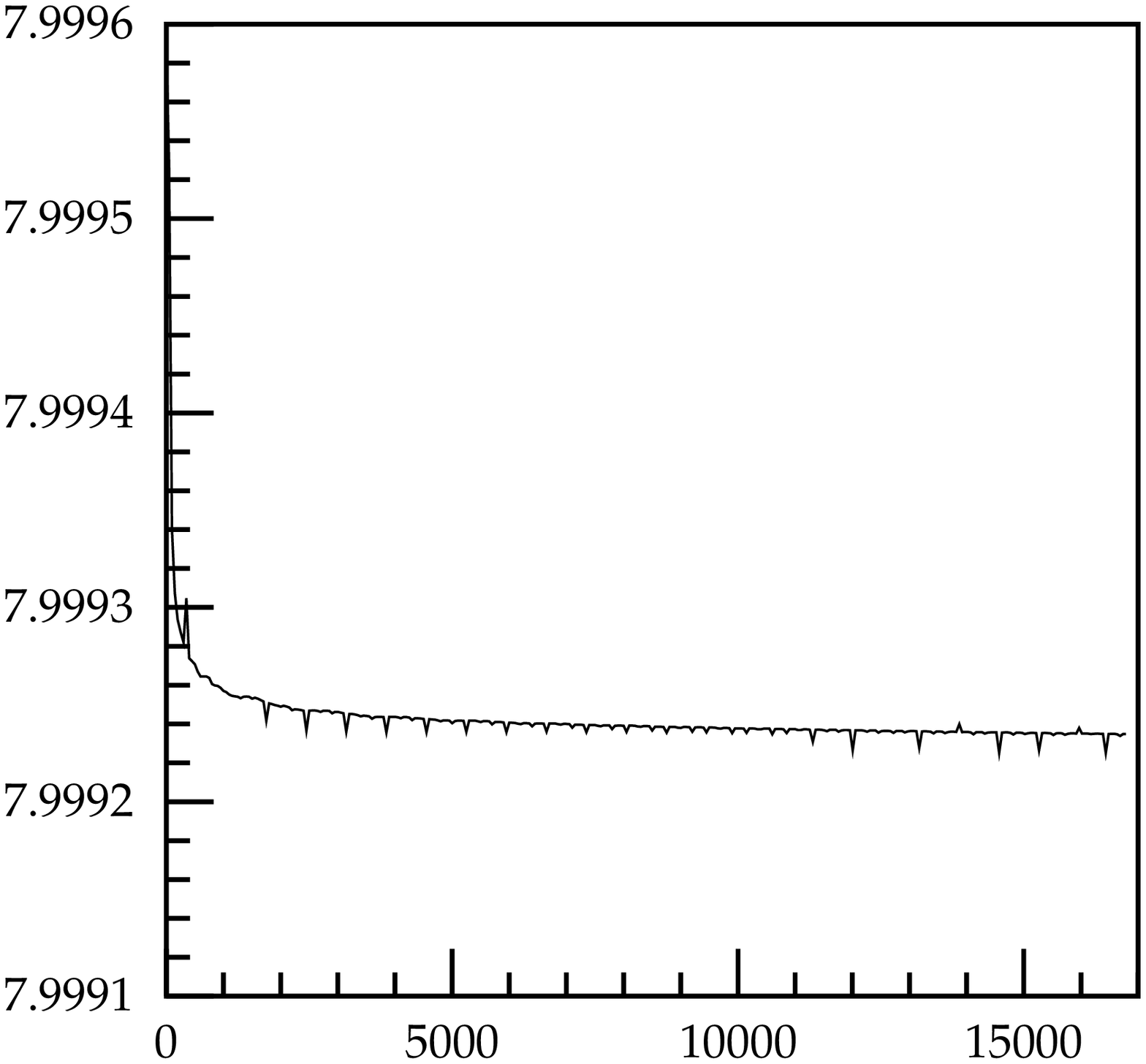}}
 \epsfxsize=7cm \put(9,0){\epsffile{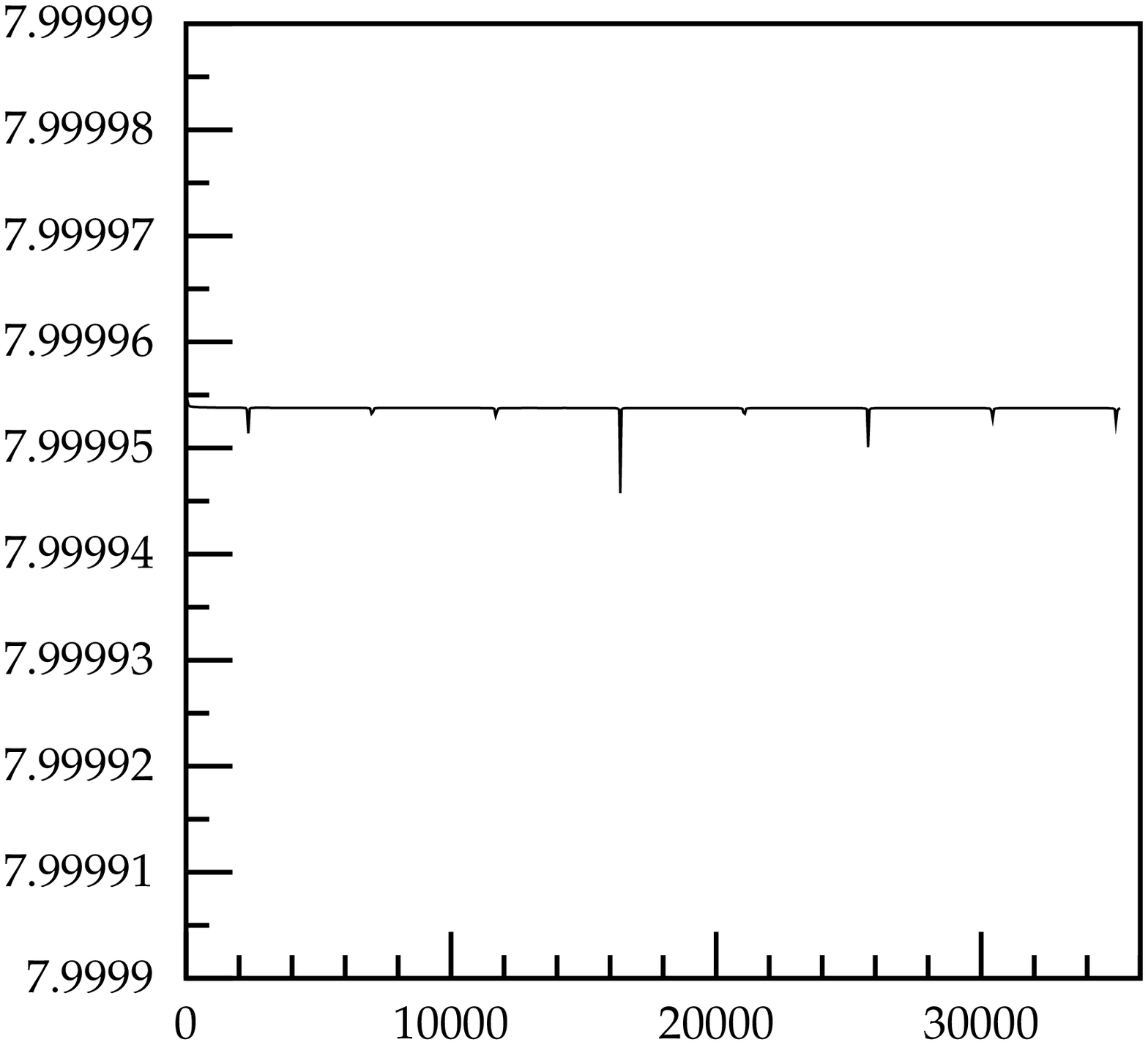}}

\put(1,0){a}
\put(7,0){b}
\put(13,0){c}

\end{picture}
\caption{\label{aaa}  Total energies of field configurations seen for $n=2$, started initially
with a kink and an antikink separated by
a) $d=3.0$, b) $d=5$, c) $d=8$. The little `blips' are the numerical artifacts of calculation
of energies when the breather is very thin.
}
\end{figure}

\begin{figure}[htbp]
\unitlength1cm \hfil
\begin{picture}(7,8)
 \epsfxsize=7cm \put(0,0){\epsffile{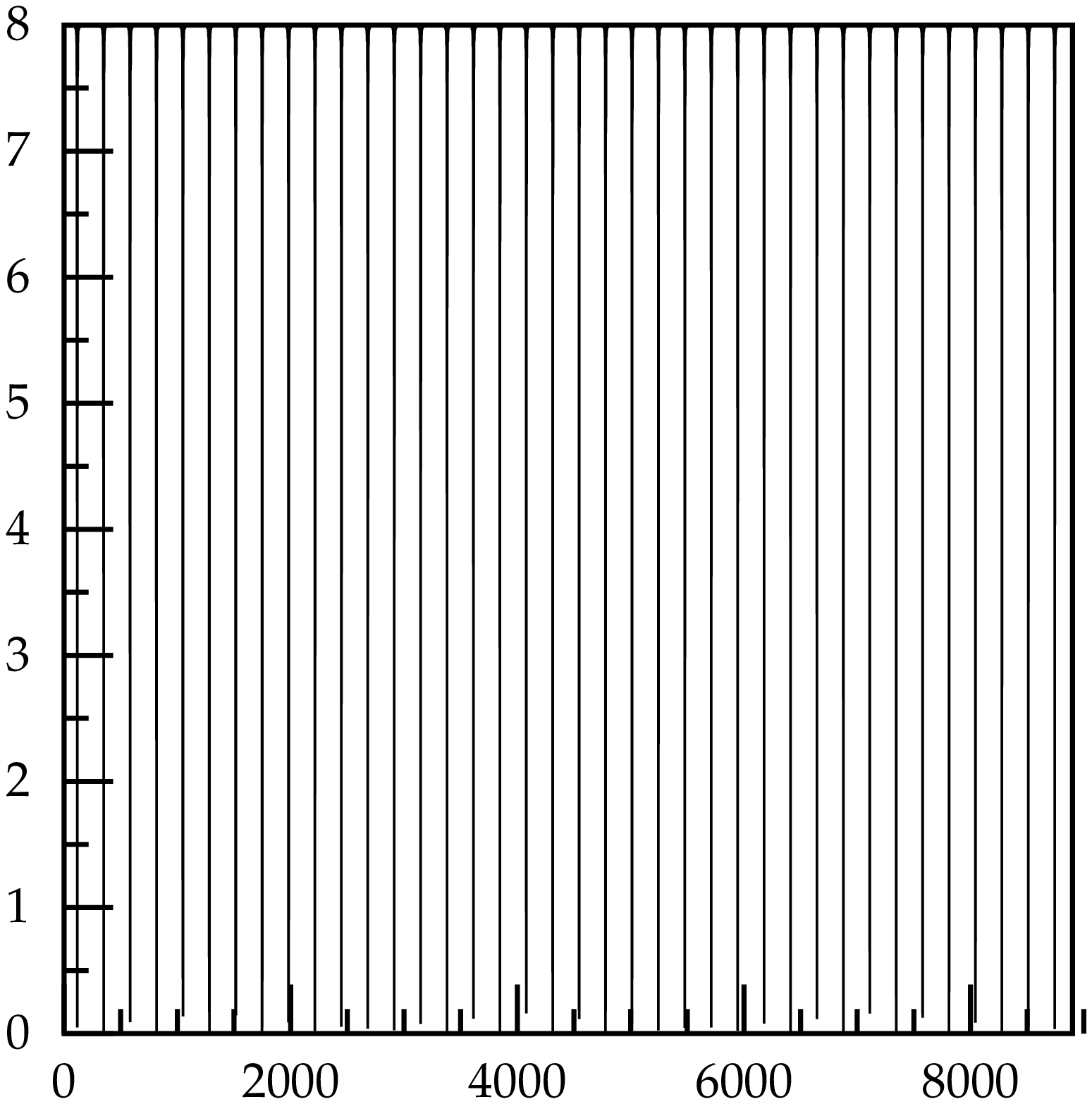}}
\end{picture}
\caption{\label{bbb} Potential energy as function of time for $n=2$ for the case of 
initial separation $d=5$
}
\end{figure}

\begin{figure}[htbp]
\unitlength1cm \hfil
\begin{picture}(7,8)
 \epsfxsize=7cm \put(0,0){\epsffile{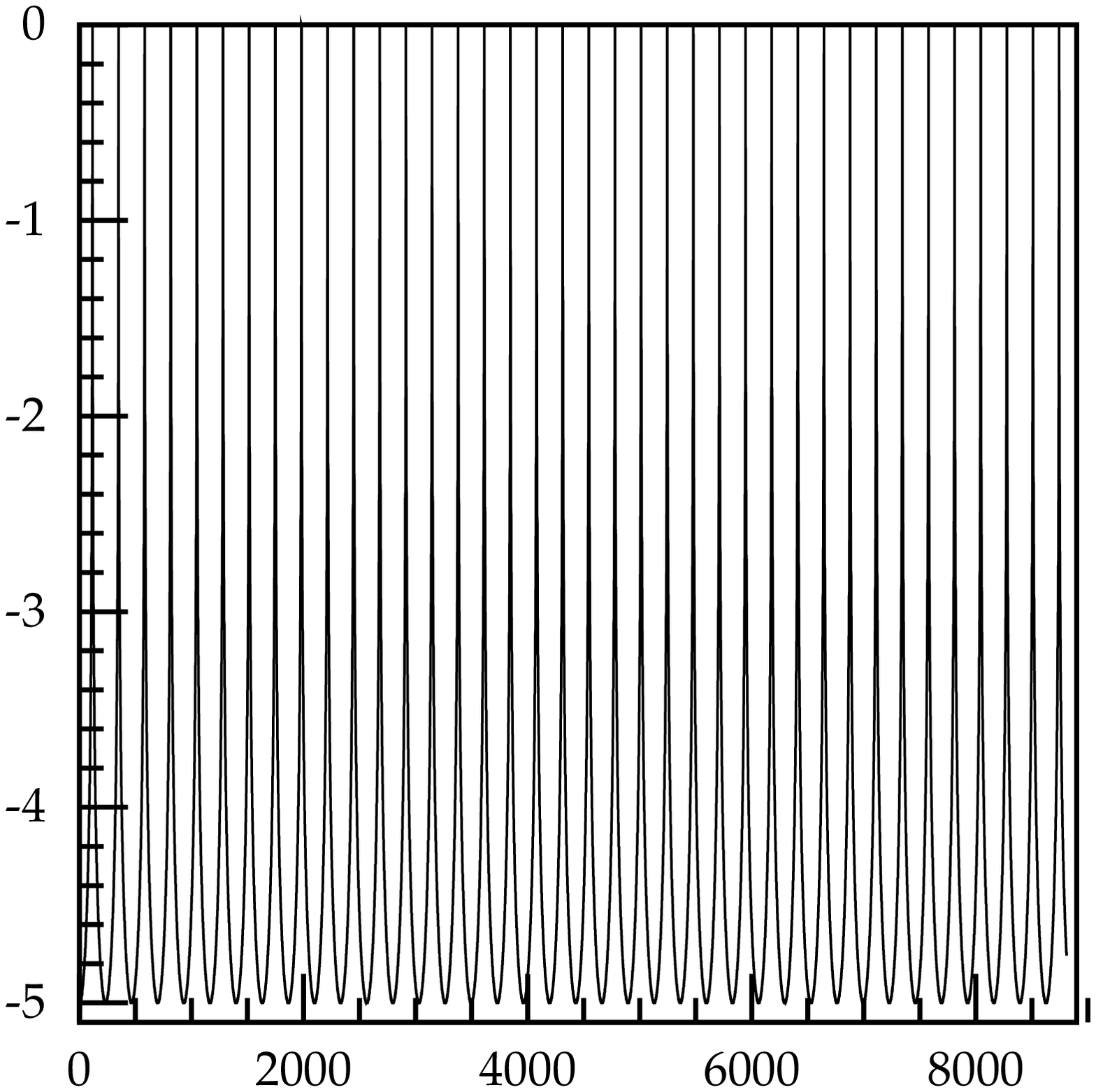}}
\end{picture}
\caption{\label{ccc} Trajectory of a kink or antikink (in $x\le 0$) for the case
of $n=2$ when the initial configuration had $d=5$.
}
\end{figure}

We see that from $t\sim 100$ the field is essentially of the breather type.
In fact - this can be checked by looking at the field itself at various times.
Looking at the plots in fig.3 we can evaluate the periods of oscillations of the 
breathers. We find that for $d=5$ this period is approximately $436.29$ giving $\omega=0.0134$, for
$d=3$ its is $T\sim 63.28$ - giving $\omega=0.094$ and for $d=4$ (not shown here) - it is $T\sim 171.61$
corresponding to $\omega\sim 0.0366$. 
As total energy is given by
\be 
E\,=\,2E_0\sqrt{1-\omega^2},
\ee
where $E_0$ is the energy of one kink (in our case normalised to $4$) we see that the corresponding 
values of energies are
\be
E_{d=5}\,\sim\,7.99927,\quad E_{d=4}\,\sim\,7.9948,\quad E_{d=3}\,\sim\,7.9605
\ee
which are in excellent agreement with our numerical results.

Hence, the obtained field configurations are indeed, those of corresponding breathers.

The same method can be used for the generalised models of Bazeia et al for other values of $n$.
In the next section we look in detail at the case of $n=1$.

\section{$n=1$ model}

In this case the potential takes the form
\be 
V(\varphi)\,=\, \frac{1}{2}\frac{\sin^2(2\varphi)}{(1+\sin\varphi)^2}
\ee
This potential, due to the factor $(1+\sin\varphi)^2$ in the denominator, clearly, has a very different behaviour for $\varphi>0$ and $\varphi<0$.
Thus if we are to have a breather-like behaviour the field will vary over a different
range of values for $\varphi>0$ and $\varphi<0$. Of course, the initial kink and antikink 
can be chosen so that $\varphi>0$.

We have performed several simulations (starting with different values of $d$).
In fig 4. we present the results of the simulations corresponding to 
$d=1,3,5$ and $9.5$ (shown in fig 4a,b,c and d, respectively).
We have found that all such configurations gradually lose energy but that this loss is very slow.
Looking in detail at the plot of fig. 4 we note that this loss depends a lot on the initial
separation ({\it ie} $d$). The energy drops slowly then almost settles and then drops again
after which it gradually settles again {\it etc}. The field itself evolves very much like a 
breather (with unequal movement `up' and `down' - for the reasons mentioned above).
However, the amplitude of this movement gradually decreases. This we can see from individual 
plots of the fields or from the time evolution of the position of the kink (or antikink) 
in $x\ge 0$. This position can be determined either by looking at the values of $x$ at which the field
takes a specific value, or even better, at which the energy density has a local maximum.

\begin{figure}[htbp]
\unitlength1cm \hfil
\begin{picture}(14,14)
 \epsfxsize=7cm \put(0,7){\epsffile{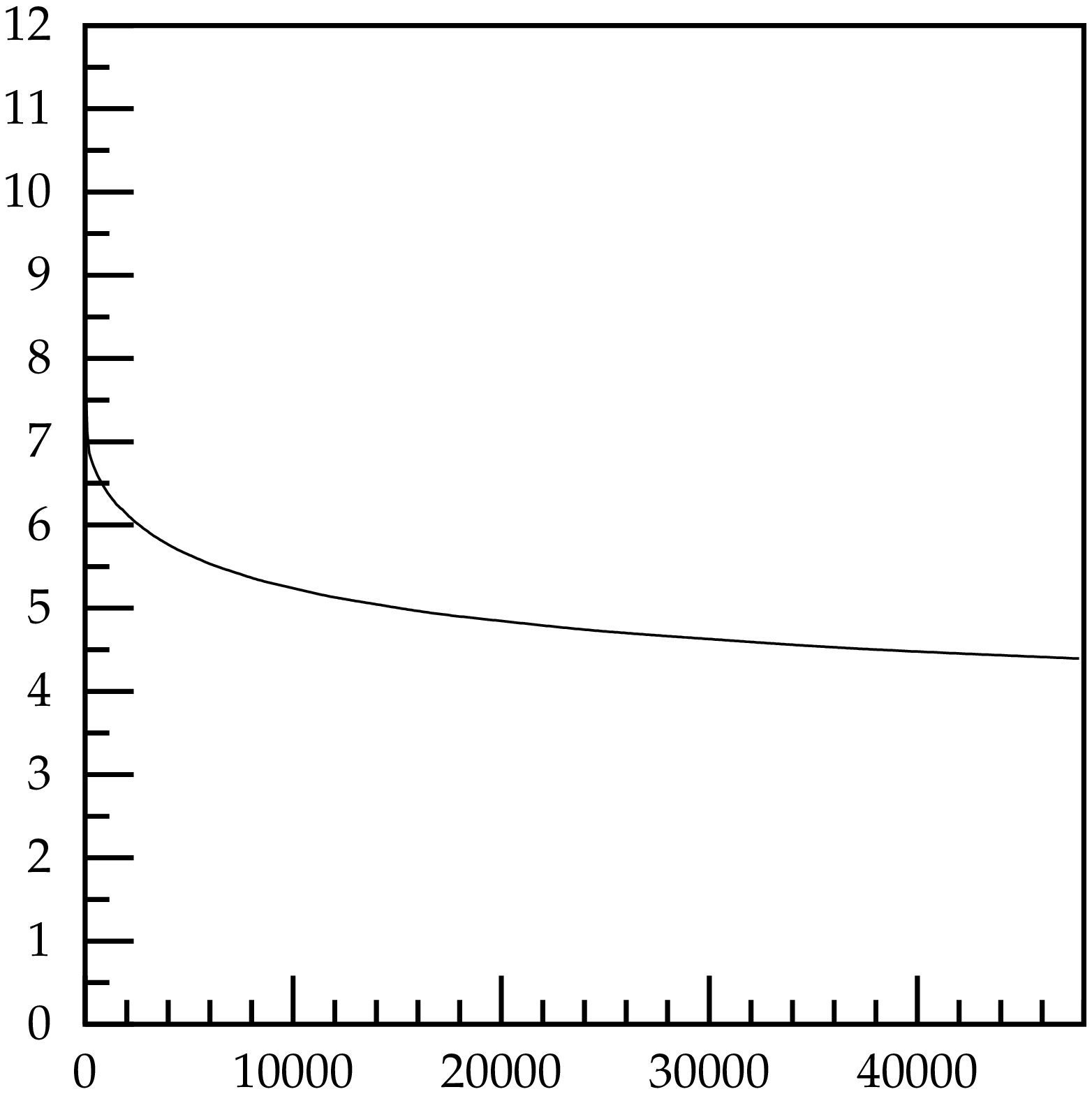}}
 \epsfxsize=7cm \put(7,7){\epsffile{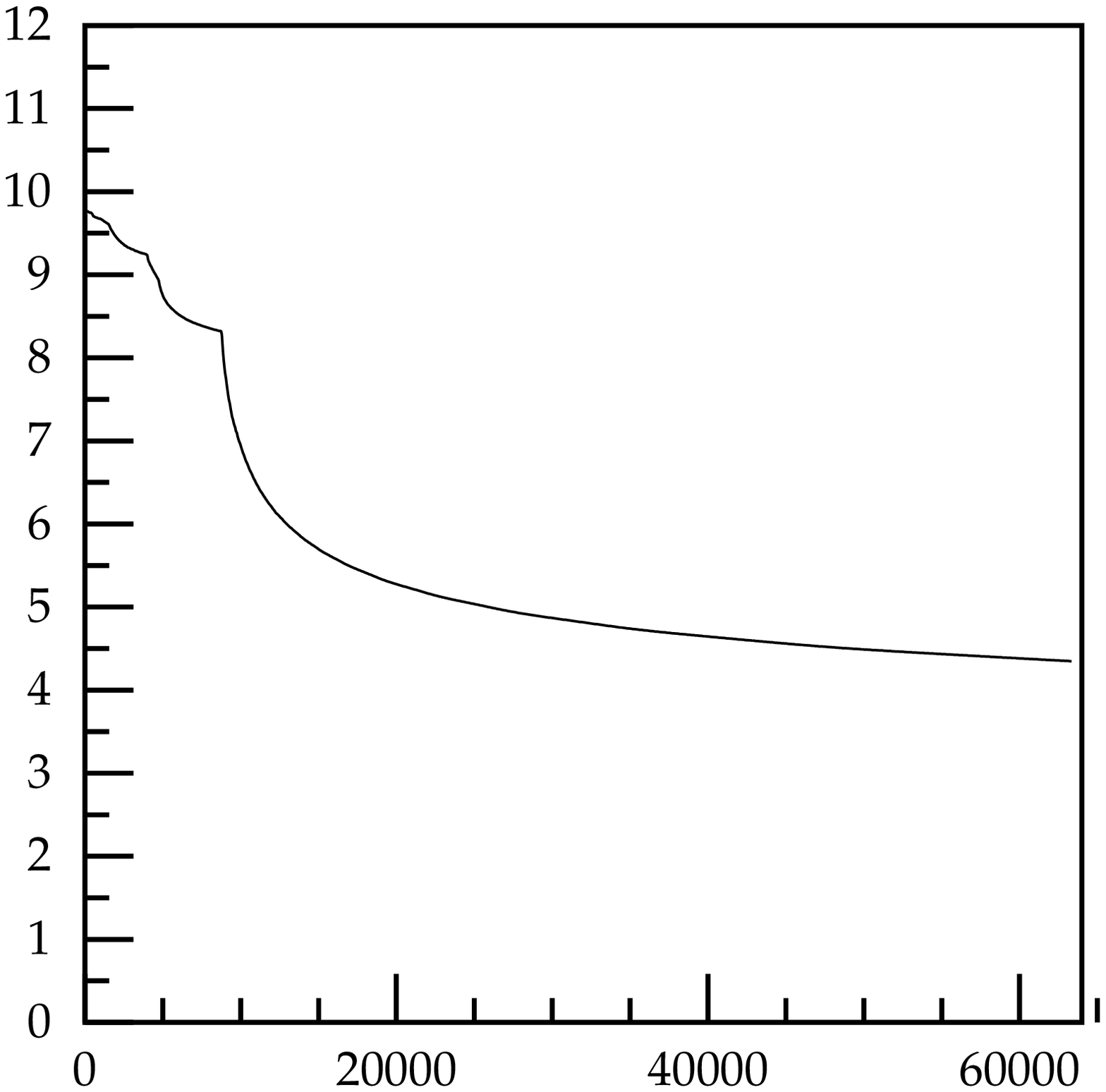}}
 \epsfxsize=7cm \put(0,0){\epsffile{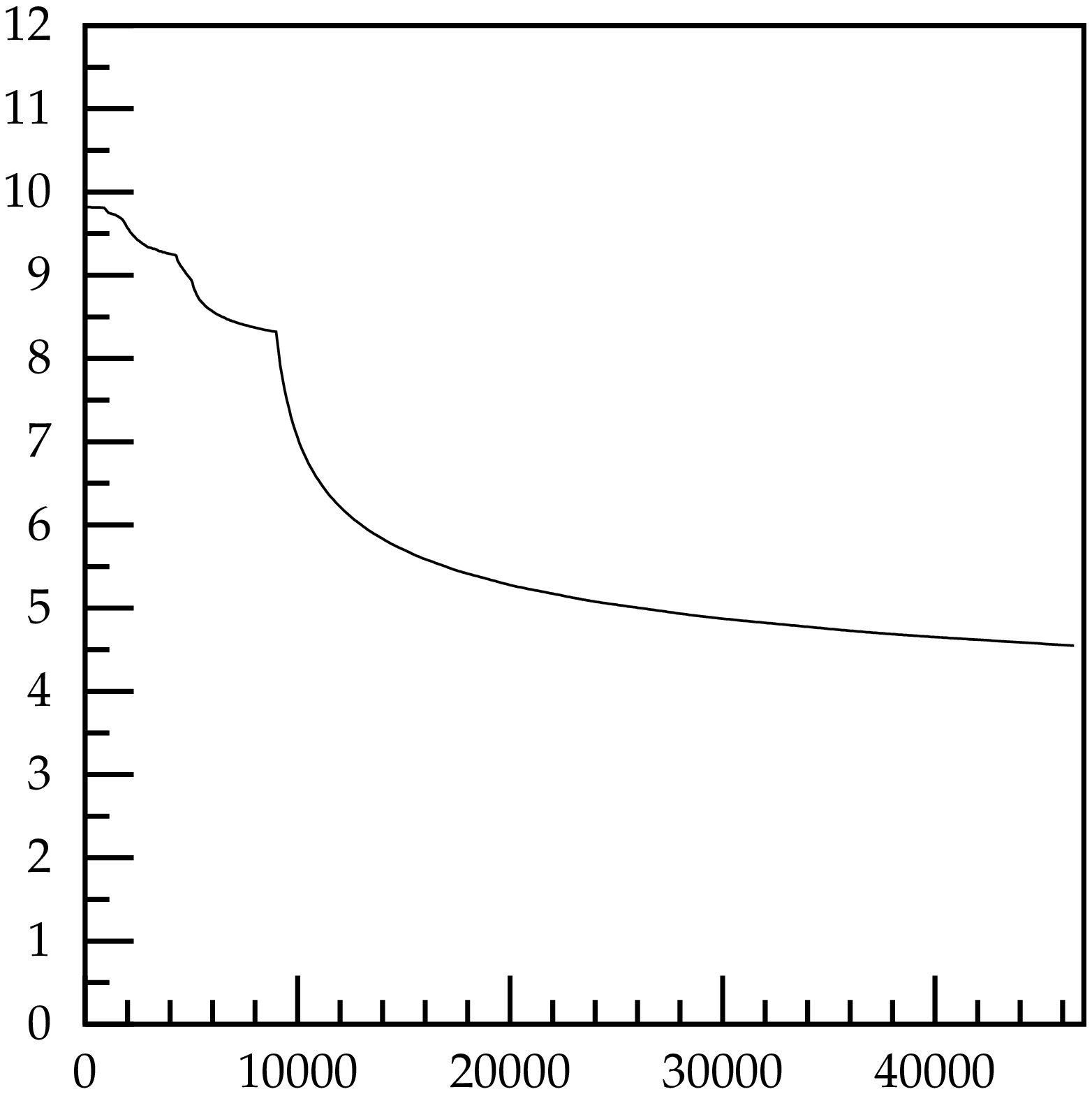}}
 \epsfxsize=7cm \put(7,0){\epsffile{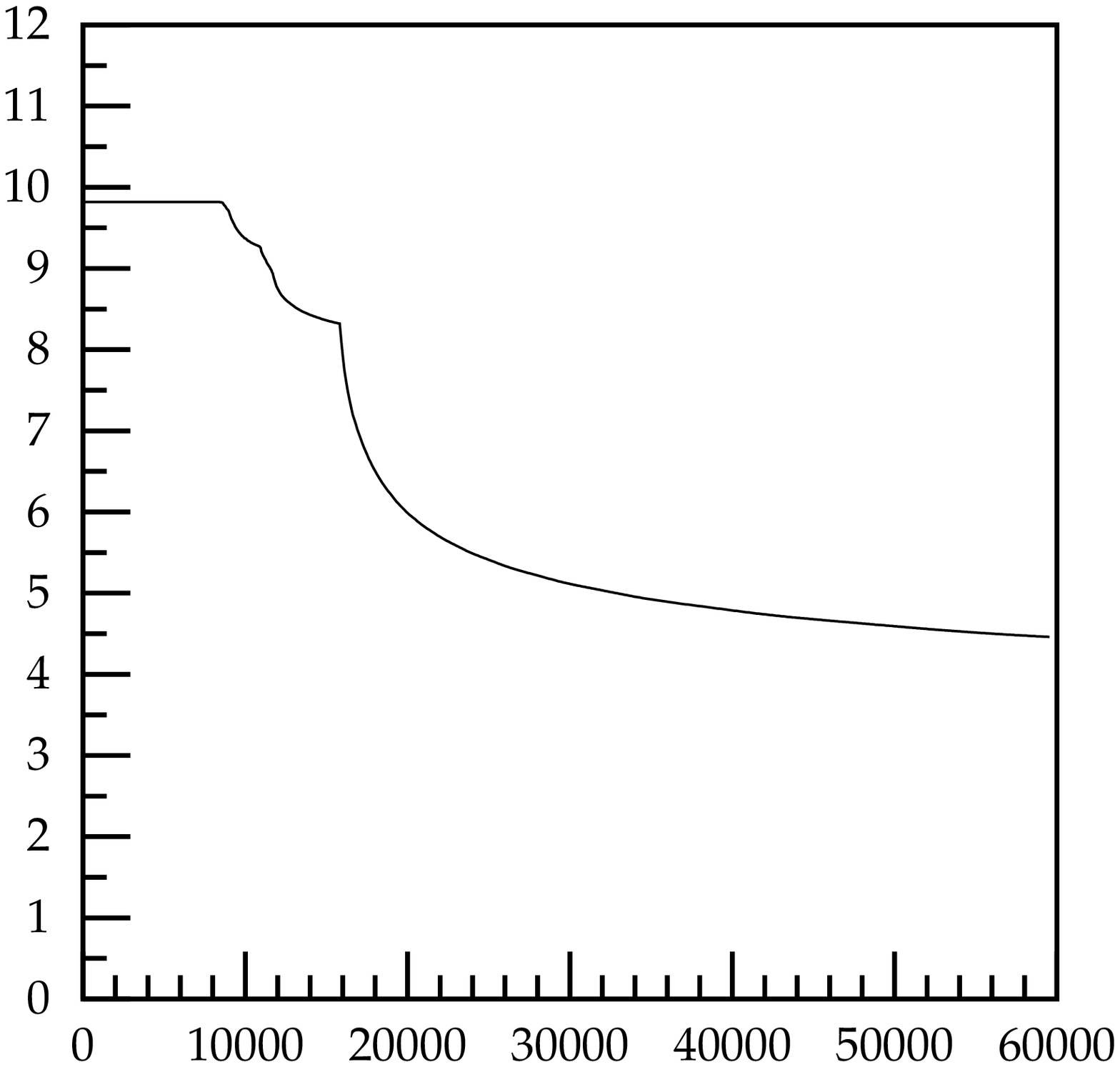}}
\put(3.5,7){a}
\put(10.5,7){b}
\put(3.5,0){c}
\put(10.5,0){d}
\end{picture}
\caption{\label{ddd} Total energies of the system for $n=1$, started with
a) $d=1$, b) $d=3$, c) $d=5$ and d) $d=9.5$.
}
\end{figure}

In fig 5. we present the plots of the time dependence of the position of the maximum 
of the energy density in $x\ge 0$. We note the above mentioned decrease of the amplitude, albeit a very slow one.
In fact the decrease gets slower and slower suggesting that even though breathers probably do not exist
the model possesses `breather-like' configurations
which are  very long lived. The actual details depend a lot on the distance between the kink and the antikink
of the initial configuration.

Our results suggest that, perhaps, the nonexistence of breathers in this model is associated with
the asymmetry of the potential as $x\leftrightarrow -x$. Hence, it is interesting to look in detail 
at the case of $n=3$ (where this effect is reduced) and at $n=4$ where the potential 
is symmetric. This we do in the next two sections.

\begin{figure}[htbp]
\unitlength1cm \hfil
\begin{picture}(14,14)
 \epsfxsize=7cm \put(0,7){\epsffile{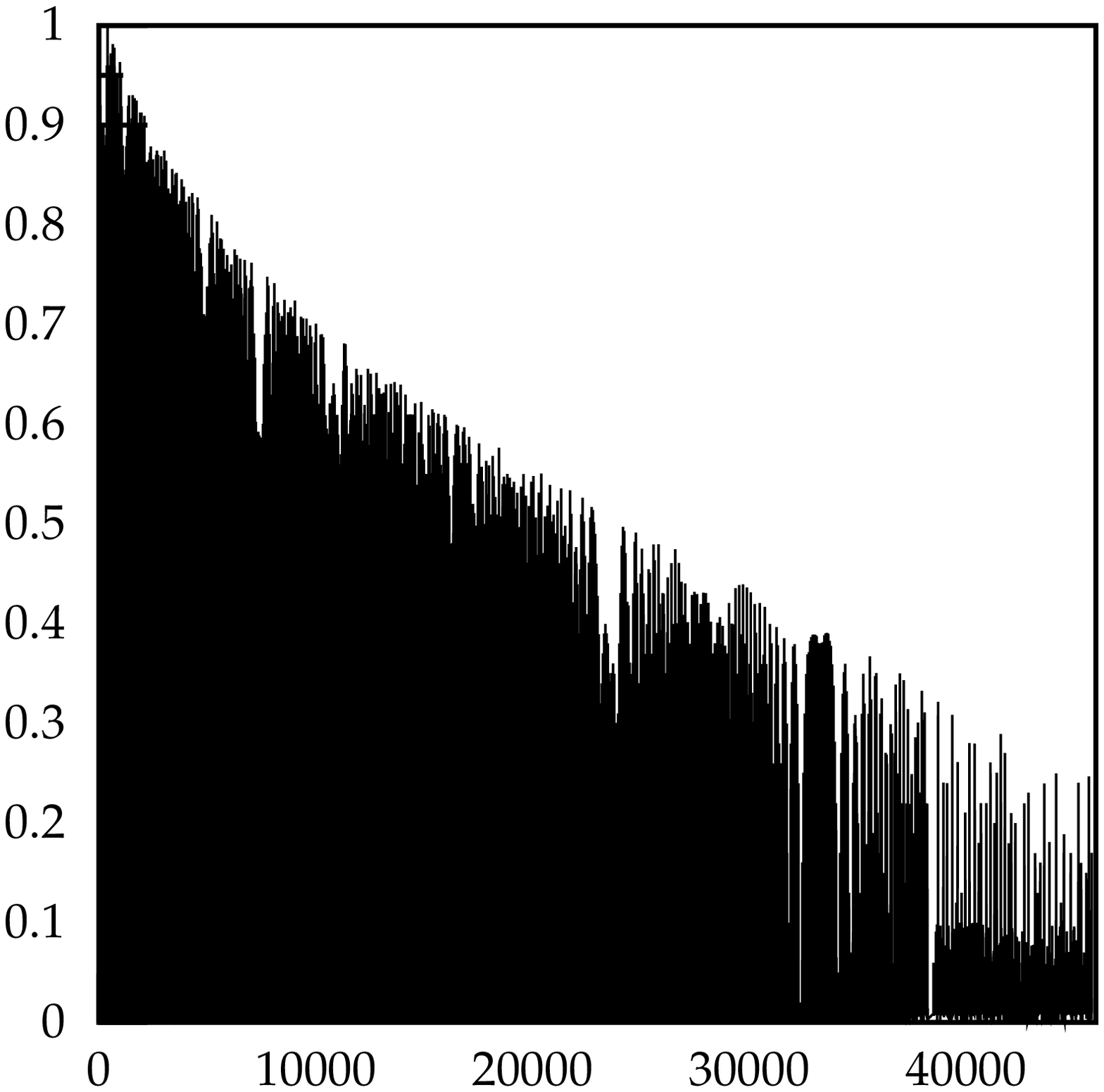}}
 \epsfxsize=7cm \put(7,7){\epsffile{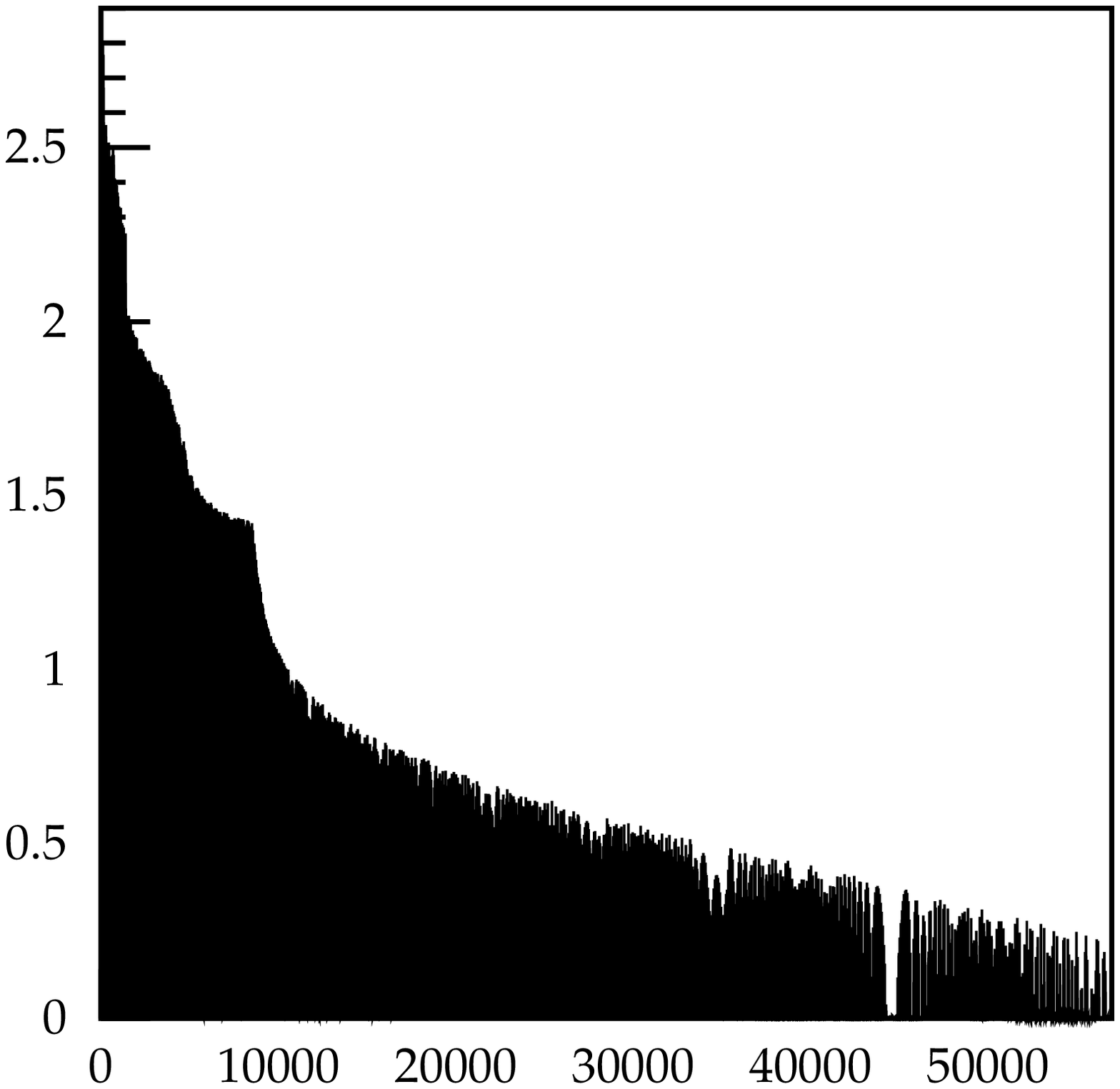}}
 \epsfxsize=7cm \put(0,0){\epsffile{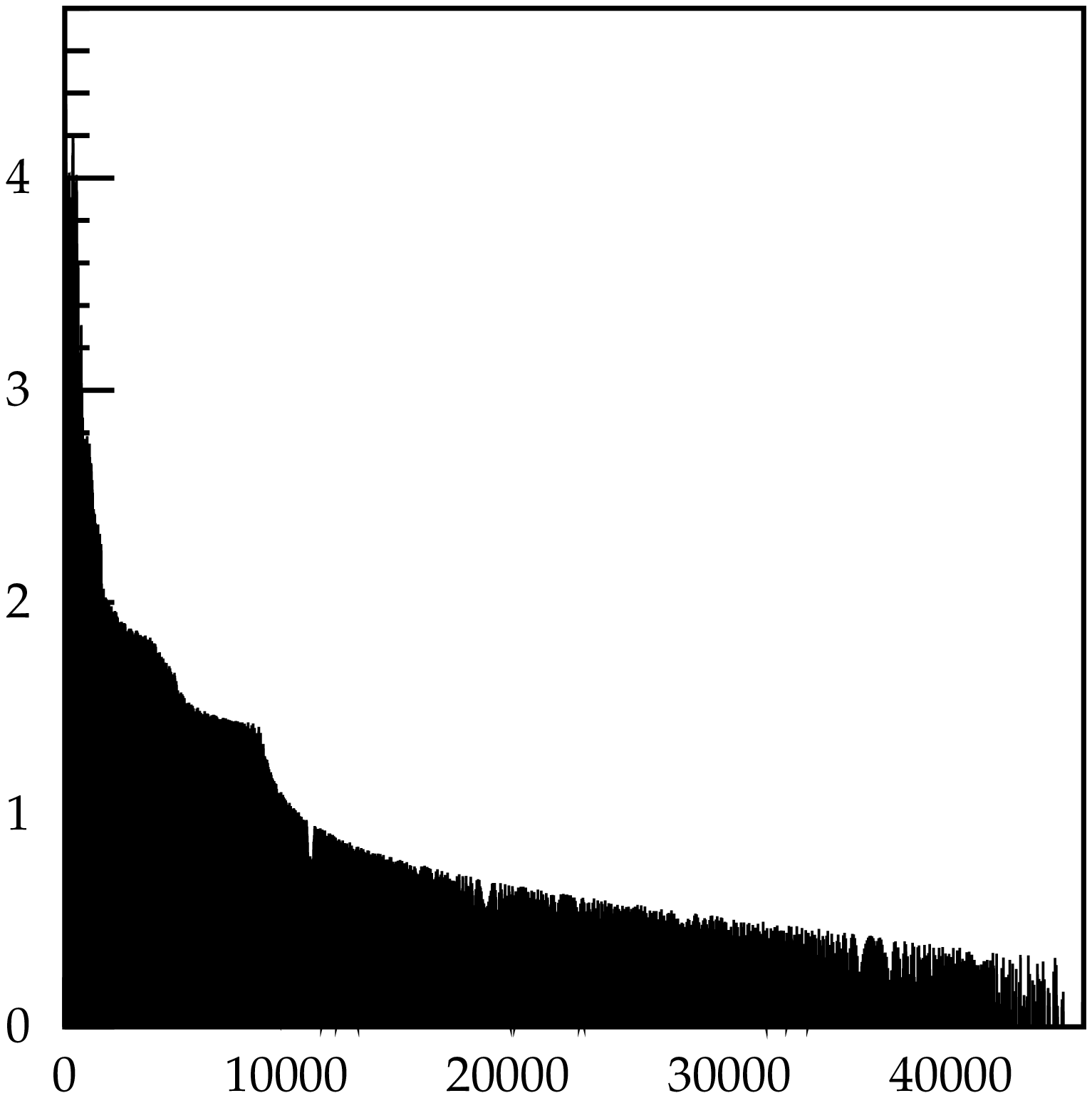}}
 \epsfxsize=7cm \put(7,0){\epsffile{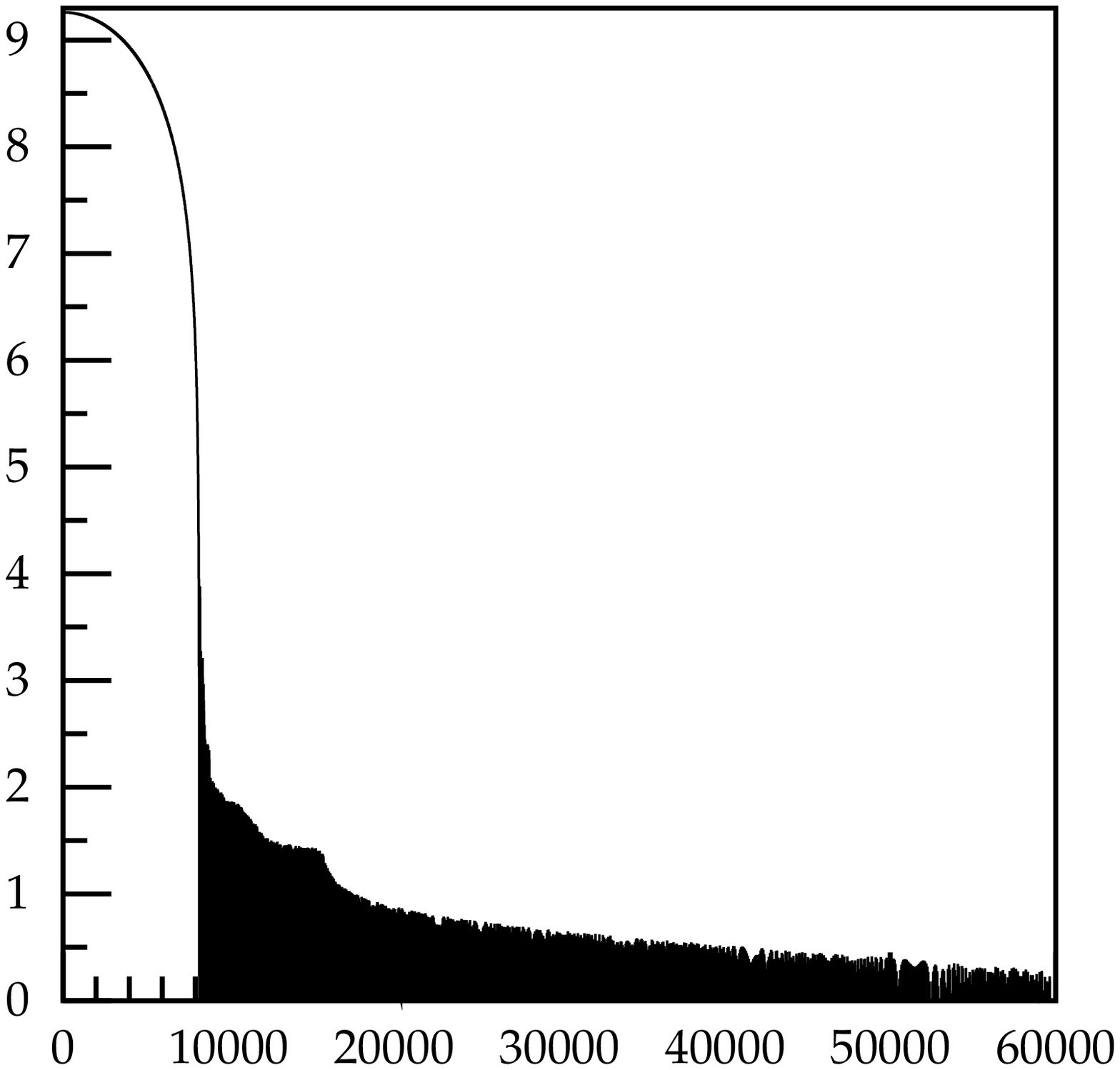}}
\put(3.5,7){a}
\put(10.5,7){b}
\put(3.5,0){c}
\put(10.5,0){d}
\end{picture}
\caption{\label{eee} Positions of the kink (or antikink) in $x\ge0$ for the simulations of 
$n=1$ case in which the initial configurations were started with
a) $d=1$, b) $d=3$, c) $d=5$ and d) $d=9.5$.
}
\end{figure}

\section{$n=3$ model}

First we consider $n=3$. This time the potential takes the form 
\be 
V(\varphi)\,=\, \frac{1}{18}\frac{\sin^2(2\varphi)}{(1+\sin\varphi)^2}(1+\sin\varphi+\sin^2\varphi)^2
\ee

Like in the previous case - the behaviour for $\varphi>0$ and $\varphi<0$ is different; however, as compared 
to the $n=1$ case the asymmetry
of the potential is somewhat reduced.

Again, we have performed many simulations for various values of $d$. In fig 6. we present  plots of the energies
for three such simulations (corresponding to $d=2,4$ and $7$). We note that, as before,  the kinks and antikinks
move towards each other and then interact with each other sending out a lot of energy (much higher percentage than in the $n=1$ case
discussed before). It seems that this time the fields annihilate quite quickly and do not form a long-living
`breather' - or if they do this breather is quite broad.
To test this last idea we have decided to rerun one of the cases considered before
, namely $d=4$, with larger $dx$ 
- so that our lattice is larger, namely $-75<x<75$; we show it as fig. 6d.
We do not see much difference - suggesting that the effects we have seen are genuine; {\it ie} that the field
decays much more quickly than in the $n=1$ case.

\begin{figure}[htbp]
\unitlength1cm \hfil
\begin{picture}(14,14)
 \epsfxsize=7cm \put(0,7){\epsffile{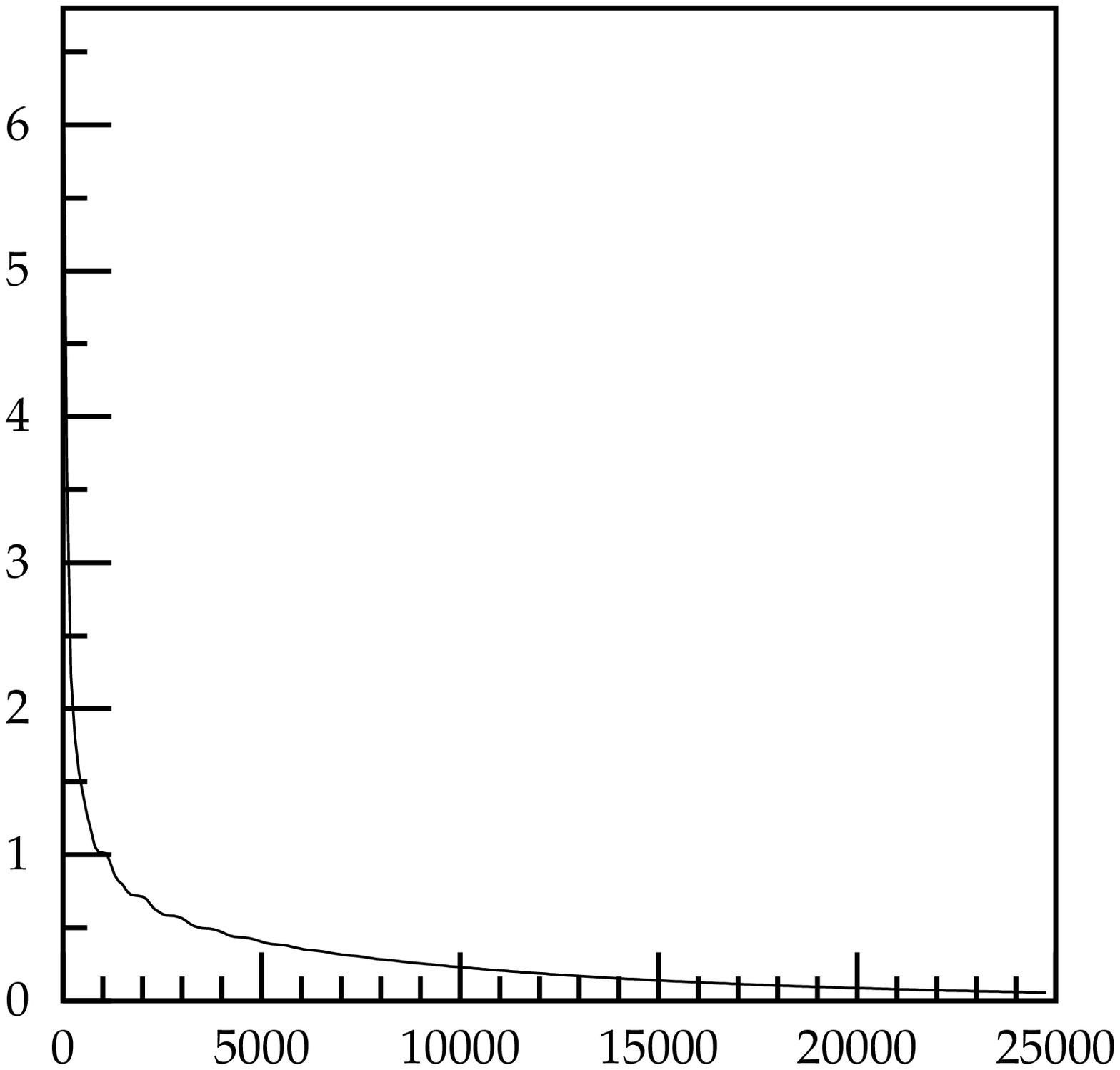}}
 \epsfxsize=7cm \put(7,7){\epsffile{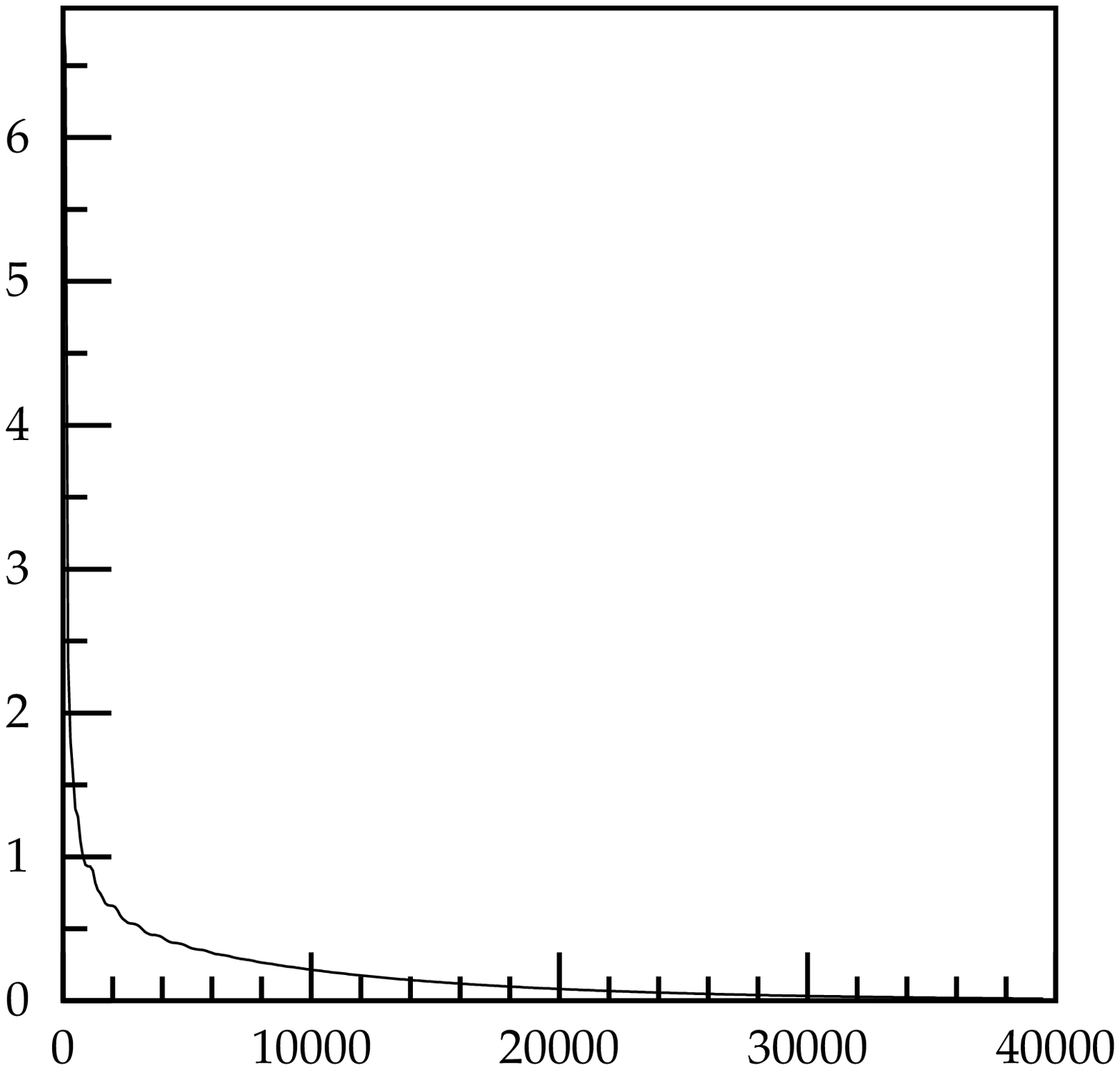}}
 \epsfxsize=7cm \put(0,0){\epsffile{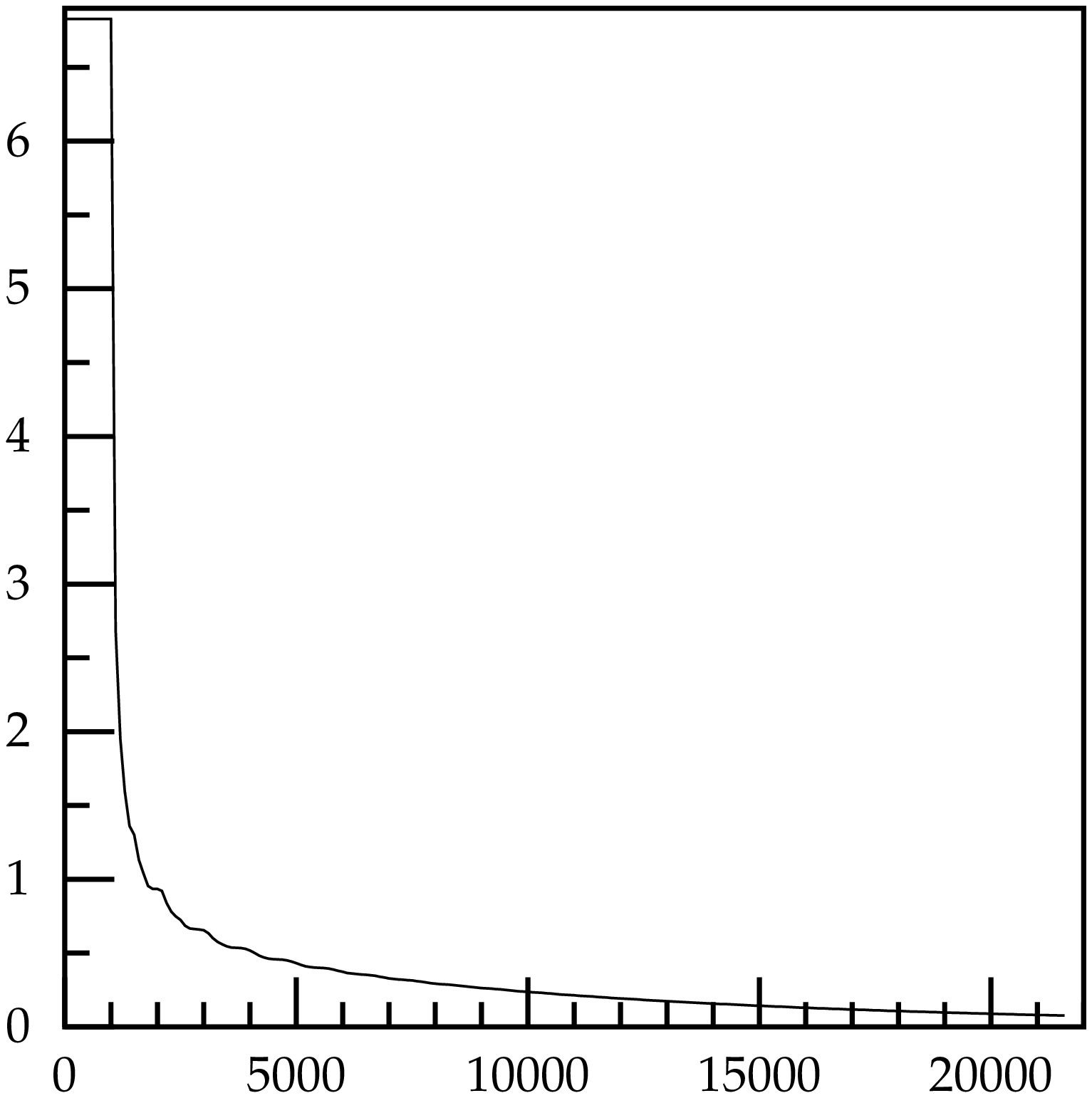}}
 \epsfxsize=7cm \put(7,0){\epsffile{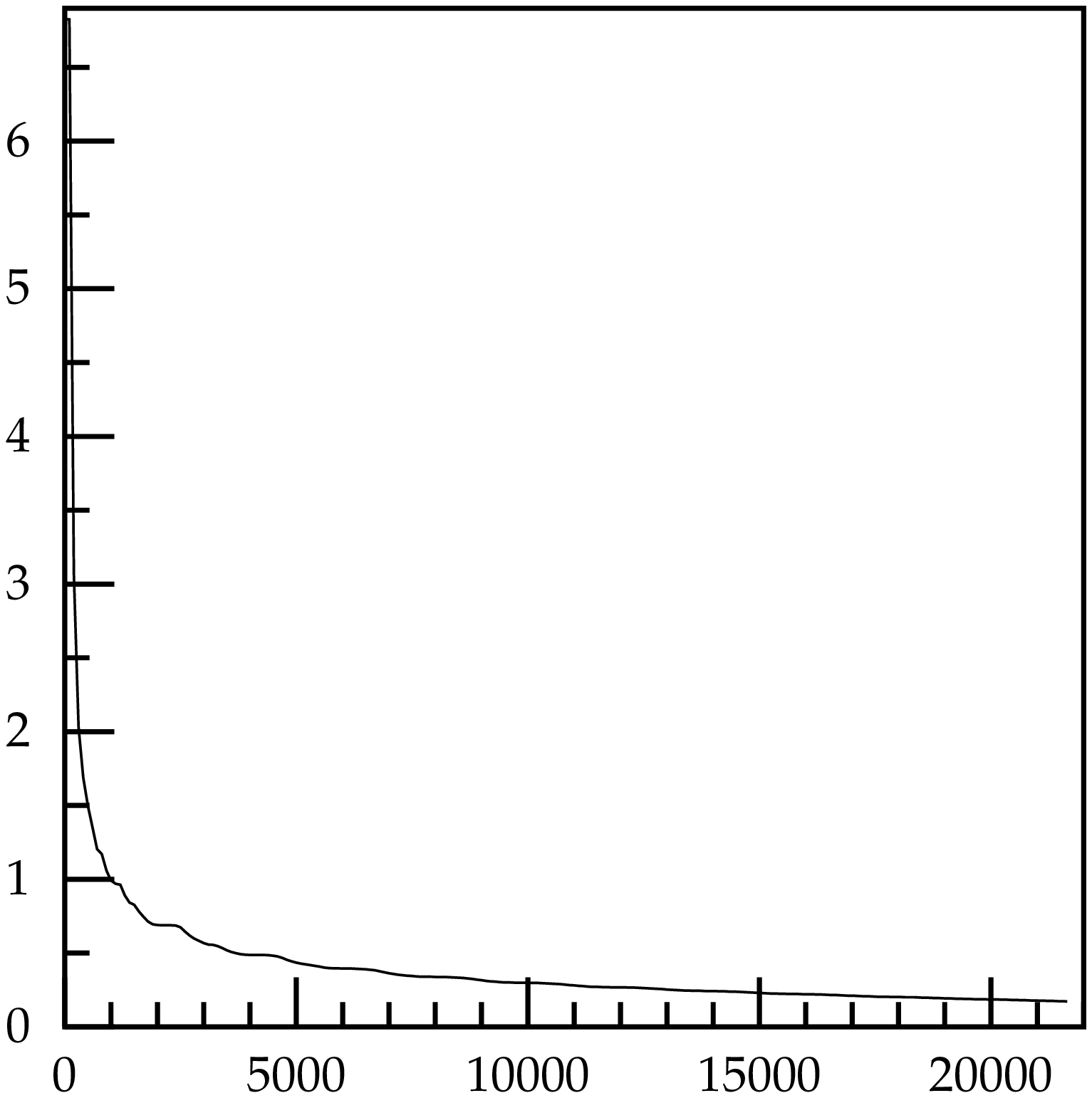}}
\put(3.5,7){a}
\put(10.5,7){b}
\put(3.5,0){c}
\put(10.5,0){d}
\end{picture}
\caption{\label{traj_L10} Total energies of the system for $n=3$, started with
a) $d=2$, b) $d=4$, c) $d=7$ and d) $d=4$ but with larger $dx$
}
\end{figure}

\section{$n=4$ model}
Finally we have also looked at the $n=4$ case. 
This time the potential is given by
\be 
V(\varphi)\,=\, \frac{1}{32}\sin^2(2\varphi)(1+\sin^2\varphi)^2
\ee
and, like the Sine Gordon case, the potential exibits symmetry $x\leftrightarrow -x$.

\begin{figure}[htbp]
\unitlength1cm \hfil
\begin{picture}(16,8)
 \epsfxsize=8cm \put(0,0){\epsffile{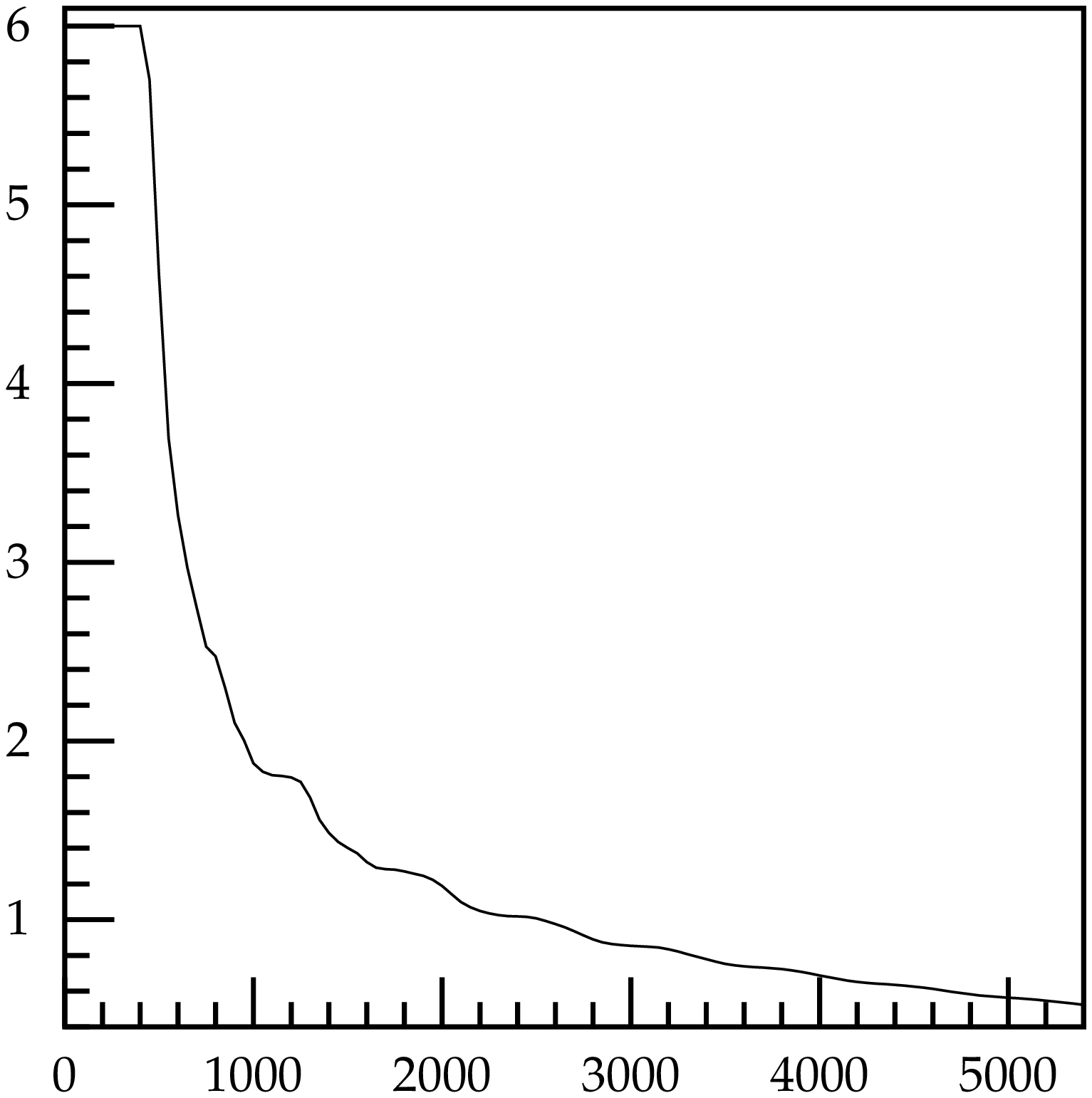}}
 \epsfxsize=8cm \put(8,0){\epsffile{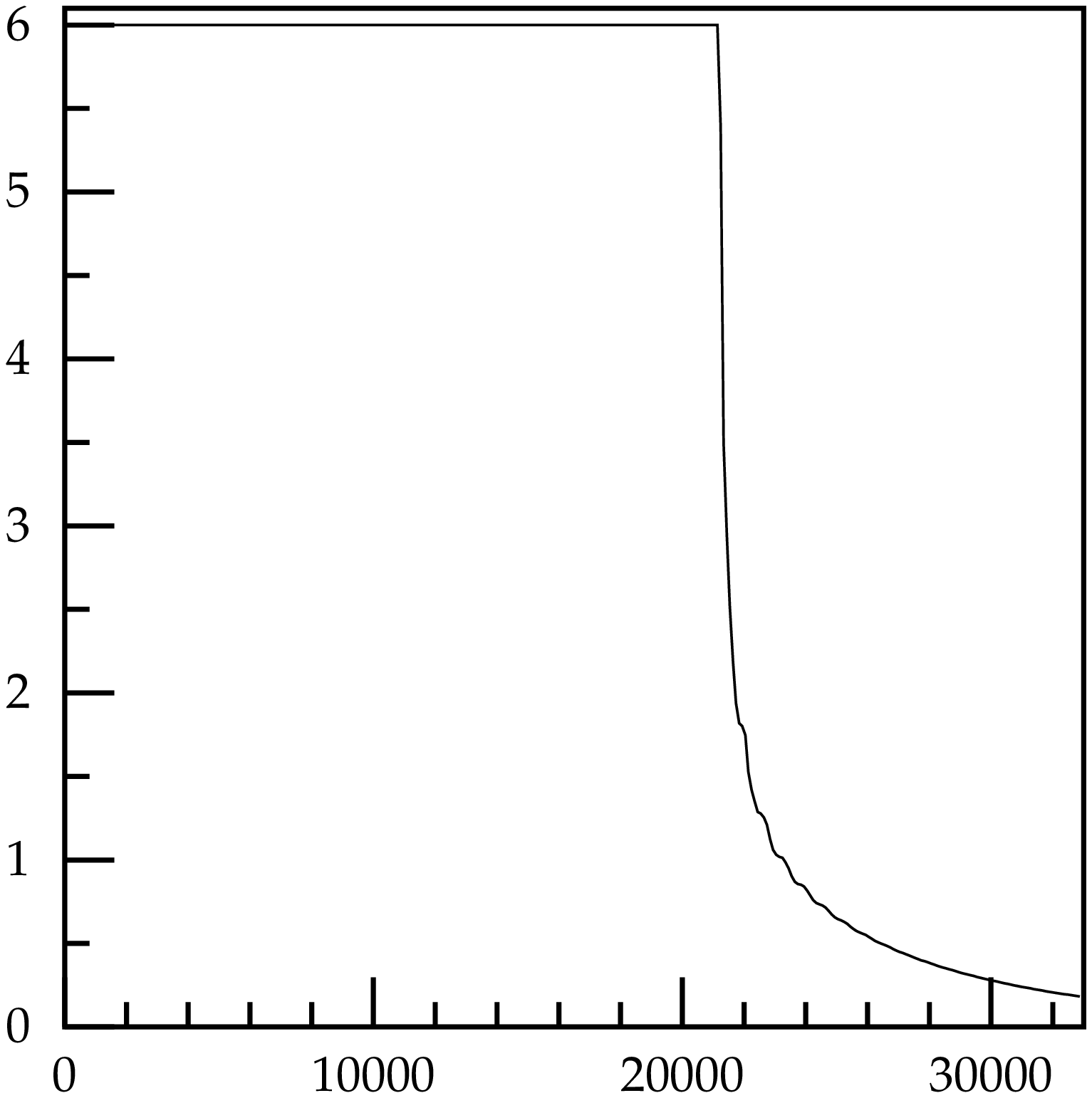}}
 \put(4,0){a}
\put(12,0){b}
\end{picture}
\caption{\label{traj_L10B} Total energy of the system started with a) $d=6$, b) $d=10$.}
\end{figure}

However, performing the simulations in this case we have quickly discovered that this symmetry does not matter;
all cases involved quick annihilation leading to pure radiation.
In fig 7. we present the plots of the total energy corresponding to the cases
of $d=6$ and $d=10$. We see a slow change of energy (when kink and antikink approach each other) followed by
a rapid annihilation. We have looked at many other cases - but they produced qualitatively the same 
results (varying, of course, in the time required for the original approach).
Hence, we feel that the $n=4$ model is unlikely to possess breather like solutions.

\section{More general $n$}

We have also considered breathers for other values of $n$, In all the cases we have studied the kink 
and antikink approached each other and then annihilated straight away
so it is clear that from all the models only the $n=2$ possesses real breathers and for $n=1$ 
the fields evolve into `breather-like' configurations which do not appear to be stable
but seem to be quasi-stable ({\it ie} are long-lived).

The results mentioned above were obtained in the conventional discretisation ({\it ie} the discretisation
in which the derivatives were represented by the symmetric differences of the fields etc).
The question then arises as to whether our results are `robust' with respect to the change of 
discretisation. To check that we can change the parameters of the discretisation and rerun our
numerical simulations or, go further and consider completely different discretisations.
Of these, perhaps the most interesting is the discretisation of Speight and Ward \cite{Ward} as it 
preserves the Bogomolnyi bound \cite{Bogomolny}. Hence in the next section we discuss such 
discretisations applied to our case.

\section{Topological discretisation}

Consider the first class of Bazeia etc all models and look at their Lagrangian (\ref{lag}) and (\ref{potone}).
Take the static part and rewrite (\ref{lag}) and (\ref{potone})
as 
\begin{equation}
L\,=\,(\partial_x\varphi)^2\,+\,\left(\frac{2\lambda}{n}\tan\varphi\,+\,\frac{2\lambda}{n}\frac{\sin^{n+1}\varphi}{\cos\varphi}\right)^2
\label{bogo}
\end{equation}
$$=\,\left(\partial_x\varphi\,+\,\frac{2\lambda}{n}\tan\varphi\,-\,\frac{2\lambda}{n}\frac{\sin^{n+1}\varphi}{\cos\varphi}\right)^2\,-\, 2\partial_x\varphi\left(\,\frac{2\lambda}{n}\tan\varphi\,+\,\frac{2\lambda}{n}\frac{\sin^{n+1}\varphi}{\cos\varphi}\right)$$
However, when we calculate the total energy we integrate all terms above over $x$  and we spot that the last
term is a total divergence as we can write
\begin{equation}
\label{total}
2\partial_x\varphi\,\left(\frac{2\lambda}{n}\tan\varphi\,+\,\frac{2\lambda}{n}\frac{\sin^{n+1}\varphi}{\cos\varphi}\right)\,=\,\partial_x F,
\end{equation}
where the explicit form of $F$ can be easily calculated for each $n$.

Hence after the integration over $x$ the term (\ref{total}) gives only the boundary contribution
$F(x=\infty)-F(x=-\infty)$.
So, in the continuum limit, we have the Bogomolny equation for $\varphi$ which is given by
\begin{equation}
\label{bogom}
\partial_x\varphi\,=-\,\left(\frac{2\lambda}{n}\tan\varphi\,+\,\frac{2\lambda}{n}\frac{\sin^{n+1}\varphi}{\cos\varphi}\right)
\end{equation}
and the energy of all fields for which $F(x=\infty)-F(x=-\infty)$ is given is smallest for the fields which satisfy
(\ref{bogom}). Thus the static kink is given by the solution of (\ref{bogom}). As this is a first order equation 
for $\varphi$ and is invariant under translations in $x$, its solution is unique, up to the constant of integration $x_0$,
which comes through $\varphi=\varphi(x-x_0)$ where $x_0$ is the position of the kink.

The discretisation of Speight and Ward \cite{Ward} is based on the observation that one can perform the discretisation
of the Lagrangian (\ref{bogo}) so that  (\ref{total}) still holds.
Notice that this is possible as the two terms on the left hand side of (\ref{total}) are the two terms that appear 
in the Langrangian (\ref{lag}) and (\ref{potone}). 
 Hence with such a discretisation the minimum field is the field which solves the discrete Bogomolny' equation
 (again involving the same two terms) and its static solution is exact on the lattice. Thus such a field
 does not experience the Peirels Nabarro potential and, as shown by Speight and Ward in their paper \cite{Ward}
 and in many subsequent papers \cite{Speight}, the numerical work with such fields has fewer numerical artifacts.
 
 So, here we follow the ideas of Speight and Ward and consider topological discretisation of (\ref{lag}) and 
 (\ref{potone}). However, a little thought shows that this is not so simple. For $n\ne2$ the expression 
 for (\ref{potone}) is quite complicated and so the problem is not very simple.
 
 Notice that our expressions for $F$ (\ref{total}) are very different for $n$ odd and $n$ even.
 For $n$ even we have (ignoring overall factors $\frac{2}{n}$)
 
\begin{equation}
n=2,\qquad\qquad  V\sim \tan \varphi -\frac{\sin^3\varphi}{\cos \varphi},\qquad \rightarrow \qquad F=\frac{1}{2}\sin{2\varphi},
\end{equation}
\begin{equation} 
n=4,\qquad V\sim \tan \varphi -\frac{\sin^5\varphi}{\cos \varphi},\qquad \rightarrow \qquad F=\frac{3}{4}\sin{2\varphi}-\frac{1}{8}\sin{4\varphi},
\end{equation}
\begin{equation}
n=6, \qquad V\sim \tan \varphi -\frac{\sin^7\varphi}{\cos \varphi},\quad \rightarrow \quad F=\frac{1}{32}\sin{6\varphi}
-\frac{1}{4}\sin{4\varphi}-\frac{29}{32}\sin{2\varphi},
\end{equation}
etc.

However for $n$ odd we have
\begin{equation}
n=1,\qquad V\sim \tan \varphi -\frac{\sin^2\varphi}{\cos \varphi},\quad \rightarrow \quad F=-\log(1+\sin\varphi)\,+\,\sin\varphi,
\end{equation}
\begin{equation} 
n=3,\qquad V\sim \tan \varphi -\frac{\sin^4\varphi}{\cos \varphi},\quad \rightarrow \quad F=
\frac{\varphi}{8}\,-\,\frac{1}{32}\sin 4\varphi,
\end{equation}
\begin{equation}
n=5, \qquad V\sim \tan \varphi -\frac{\sin^6\varphi}{\cos \varphi},\quad \rightarrow \quad F=
\frac{6\varphi}{32}\,+\,\frac{1}{192}\sin6\varphi\,-\,\frac{3}{64}\sin4\varphi\,-\,\frac{1}{64}\sin2\varphi,
\end{equation}
etc.

Hence for $n$ even we have (relatively) simple expressions for $F$ involving sums of $\sin{2k\varphi}$ where $k$ takes  values up $k=n$ while for $n$ odd we have in addition $\log(\varphi)$.

For $n=2$ Speight and Ward took (for $x=i\,dx$)
\begin{equation}
\partial_x \varphi \quad \rightarrow\quad \left[\sin((\varphi(i+1)-\varphi(i-1))\right]\frac{1}{2\,dx}
\end{equation} and
\begin{equation}
\sin(2\varphi) \quad \rightarrow\quad \sin(\varphi(i+1)+\varphi(i-1))
\end{equation} 

Both expressions have correct limits and  the Bogomolny equation is satisfied as 
\begin{equation}
\sin\left[\varphi(i+1)+\varphi(i-1)\right] \, \sin\left[\varphi(i+1)-\varphi(i-1)\right]\,=\, \frac{1}{2}\left[\cos(2\varphi(i+1))
-\cos(2\varphi(i-1))\right].
\end{equation}
Hence, when we integrate over $x$ ({\it ie} sum over $i$), all the terms cancel except for the boundary terms.

We can repeat this procedure for, say, $n=4$ by taking again
\begin{equation}
\partial_x \varphi \quad \rightarrow\quad \left[\sin(\varphi(i+1)-\varphi(i-1))\right]\frac{1}{2\,dx}.
\end{equation}
For the terms in the potential we can take

\begin{equation}
\sin(2\varphi) \quad \rightarrow\quad \sin(\varphi(i+1)+\varphi(i-1))
\end{equation} 
and
\begin{equation}
\sin(4\varphi) \quad \rightarrow\quad \sin[\varphi(i+1)+2\varphi(i)+\varphi(i-1)].
\end{equation} 

Then the trigonometric identities mentioned above show that that this discretisation is again topological
(we have the cancelation of terms involving $\cos(2\varphi(i+1))$ and $\cos(2\varphi(i-1))$ as before and also of
$\cos[2(\varphi(i+1)+\varphi(i)]$ and $\cos[2(\varphi(i)+\varphi(i-1)]$).

In the $n=2$ case studied by Speight and Ward the Bogomolny equation (\ref{bogom}) involves just two $\sin$ functions
and is easy to solve. In our case ({\it ie} for $n=4$) the equation involves three $\sin$ functions (of 3 different
arguments) and so is much harder to solve. This procedure is even harder to apply to the case of $n=6$ where 
we have even more terms. And it seems almost impossible to generalise to the $n$ = odd cases. And the lowest
of these cases {\it ie} $n=1$ and $n=3$ are the most interesting from our point of view.

Hence we have to look at the discretisation problem differently.

\subsection{A modified approach}

Let us look at the case of $n=1$. In this case the potential is proportional to 
\begin{equation}
V\,\sim \, \tan\varphi \,-\, \frac{\sin^2\varphi}{\cos\varphi}.
\label{limi}
\end{equation}
Naive integration, mentioned above, gives us 
\begin{equation}
F\,=\, -\log(\cos\varphi)\,-\,2 Arcth(\tan\frac{\varphi}{2})\,+\,\sin\varphi.
\end{equation}

However, a few trigonometric identities tell us that $F$ is also given by
\begin{equation}
F\,=\,\sin\varphi\,-\, \log(1+\sin\varphi).
\end{equation}

This is simpler but, at first sight, does not help us in our discretisation procedure.
However, as we want to have the discrete terms in the Bogomolny bound cancel it is clear we 
would be happy if our discretisation gave us 
\begin{equation}
F\,\rightarrow\quad \sin(\varphi(i+1))-\sin(\varphi(i-1))\,-\, \log(1+\sin[\varphi(i+1)])+\log(1+\sin[\varphi(i-1)])
\end{equation}

This suggests to us that for our discrete potential we can take
\begin{equation}
V(\varphi(i+1), \varphi(i-1))\,=\,{2\,dx}\frac{F}{ \sin[\varphi(i+1)-\varphi(i-1)]}.
\end{equation}
It is easy to check that this expression has the correct continuum limit giving (\ref{limi}).

Hence our discrete Bogomolny equation becomes
\begin{equation}
K(\left[\sin((\varphi(i+1)-\varphi(i-1))\right])^2\,=\,\sin(\varphi(i+1))-\sin(\varphi(i-1))\,-\, \log(1+\sin[\varphi(i+1)])+\log(1+\sin[\varphi(i-1)]).
\end{equation}

%{\bf Can we solve this equation??}

We did not succeed to solve this equation analytically and solved it numerically.
Then we used it to construct the initial configuration which we then 
evolved using the discrete equations involving the expressions above.
Unfortunately, they reproduced the results obtained before (using the usual 
discretisation).

\section{Final Comments and Further Work}

In a way our results are disappointing; the only time when we generated breather solutions
corresponds to the Sine Gordon case, where these results are well known.
In all other cases the initial configurations involving one kink and one anti-kink
attracted, came towards each other and then annihilated. Of all our cases the $n=1$ 
model was the nearest to possessing breathers; in its case the annihilation was quite slow
and we have found that the model possesses 'breather like' long living states.
In fact, such states have lived amazing long times; as is clear from fig 5, even after
thousands of oscillations the configurations continue resembling a breather.

Of course, the nonexistence of breathers in all models apart from $n=2$ is likely
to be related to the integrability of the Sine Gordon model, and non-integrability
of the models for $n\ne2$. On the other hand the fact that the $n=1$ model
has long-living states that resemble breathers suggests that this model
is not far off from being integrable, {\it ie} is `almost integrable'.
We have no definition of this concept and we currently thinking of how to define it.

\begin{figure}[htbp]
\unitlength1cm \hfil
\begin{picture}(16,8)
 \epsfxsize=8cm \put(0,0){\epsffile{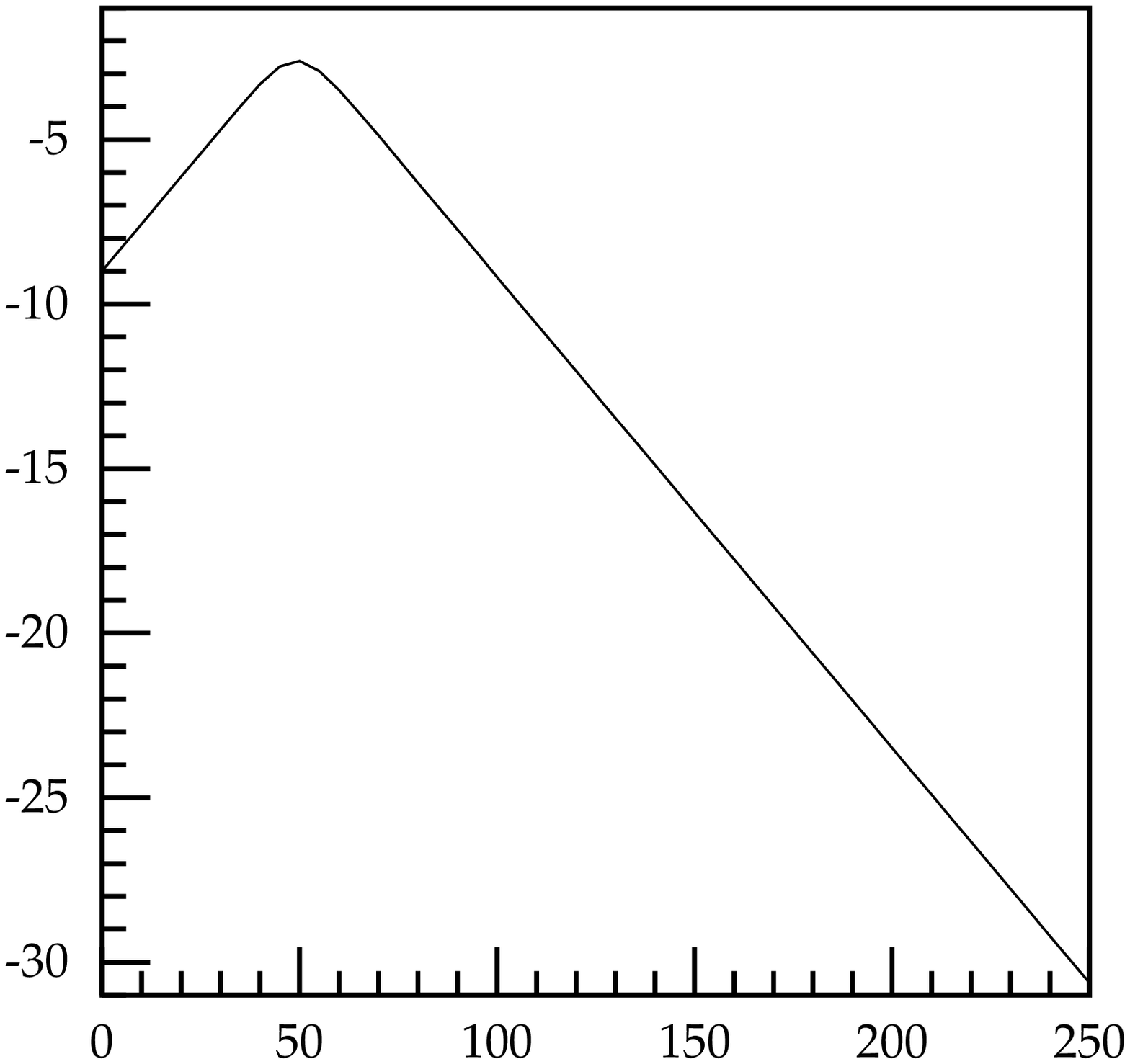}}
 \epsfxsize=8cm \put(8,0){\epsffile{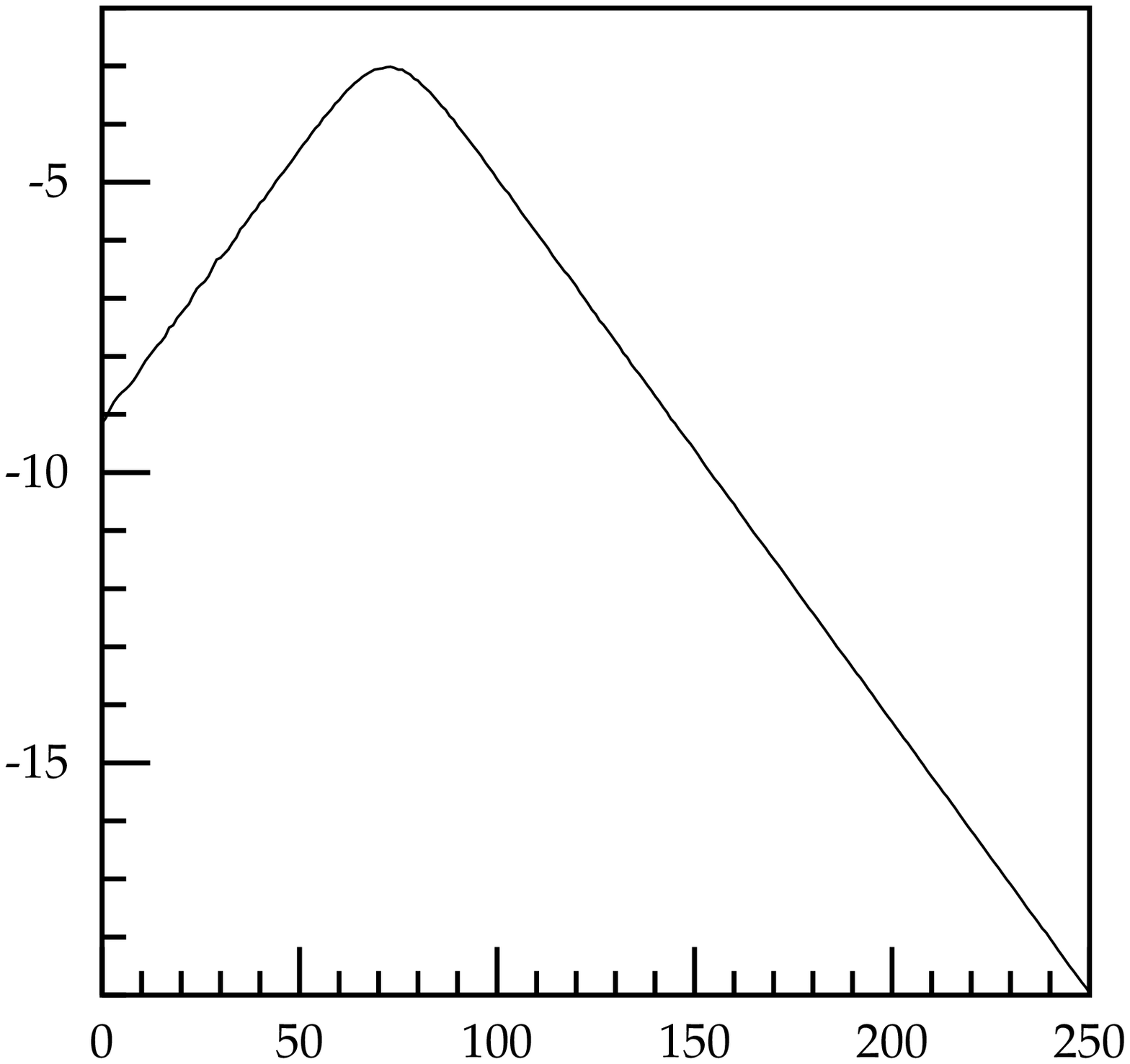}}
 \put(4,0){a}
\put(12,0){b}
\end{picture}
\caption{\label{traj_L10C} Trajectories of two kinks started with $d=9$ and with velocities $v=0.5$ in two models of even $n$: a) $n=2$, b) $n=4$.}
\end{figure}

We have also looked at two kink configurations for even $n$ (at odd $n$ such configurations
would have infinite energy). 
In the Sine-Gordon models kinks repel (although some people
argues that they go through each other). The preliminary results show that the same is true for other
values of even $n$. In fig. 8 we present the trajectories obtained by us in two simulations
(both started at $d=9$ and with the same velocity $v=0.5$. In fig 8a we present the trajectory
for the Sine Gordon case $n=2$ and in fig 8b - for $n=4$. We see very little difference
although the repulsion is stronger in the $n=4$ case. We observed the same results
for other values of $v$ and for different values of $d$. In fig 9 we show similar results
for the case of $v=0$ (this time started at a smaller $d$ - at $d=9$ but the repulsion is much 
weaker). Clearly we see repulsion in both cases. In our simulations we took
the initial conditions as the superpositions of two kinks with velocity $\pm v$ in their centre 
of mass.
\begin{figure}[htbp]
\unitlength1cm \hfil
\begin{picture}(16,8)
 \epsfxsize=8cm \put(0,0){\epsffile{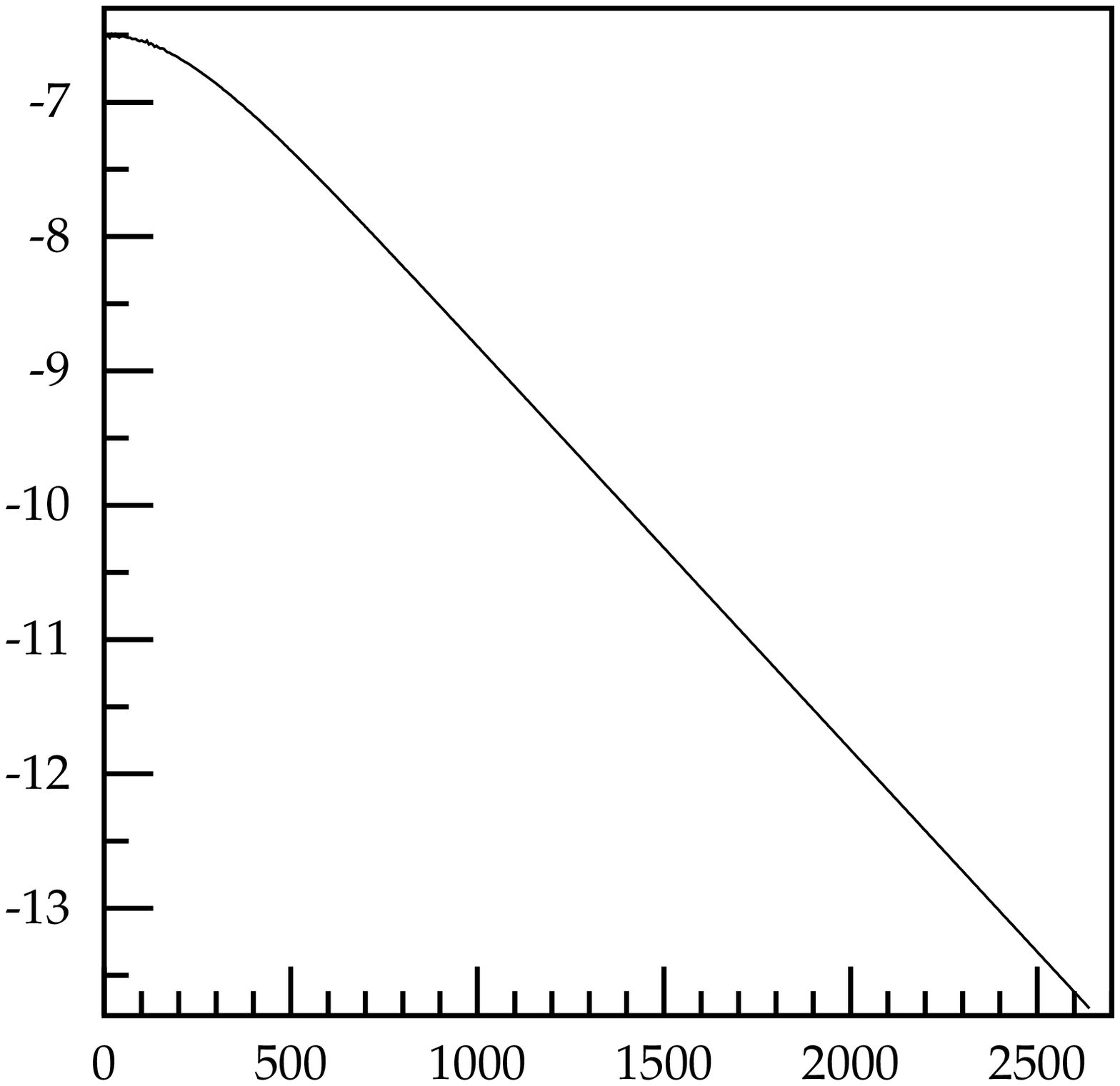}}
 \epsfxsize=8cm \put(8,0){\epsffile{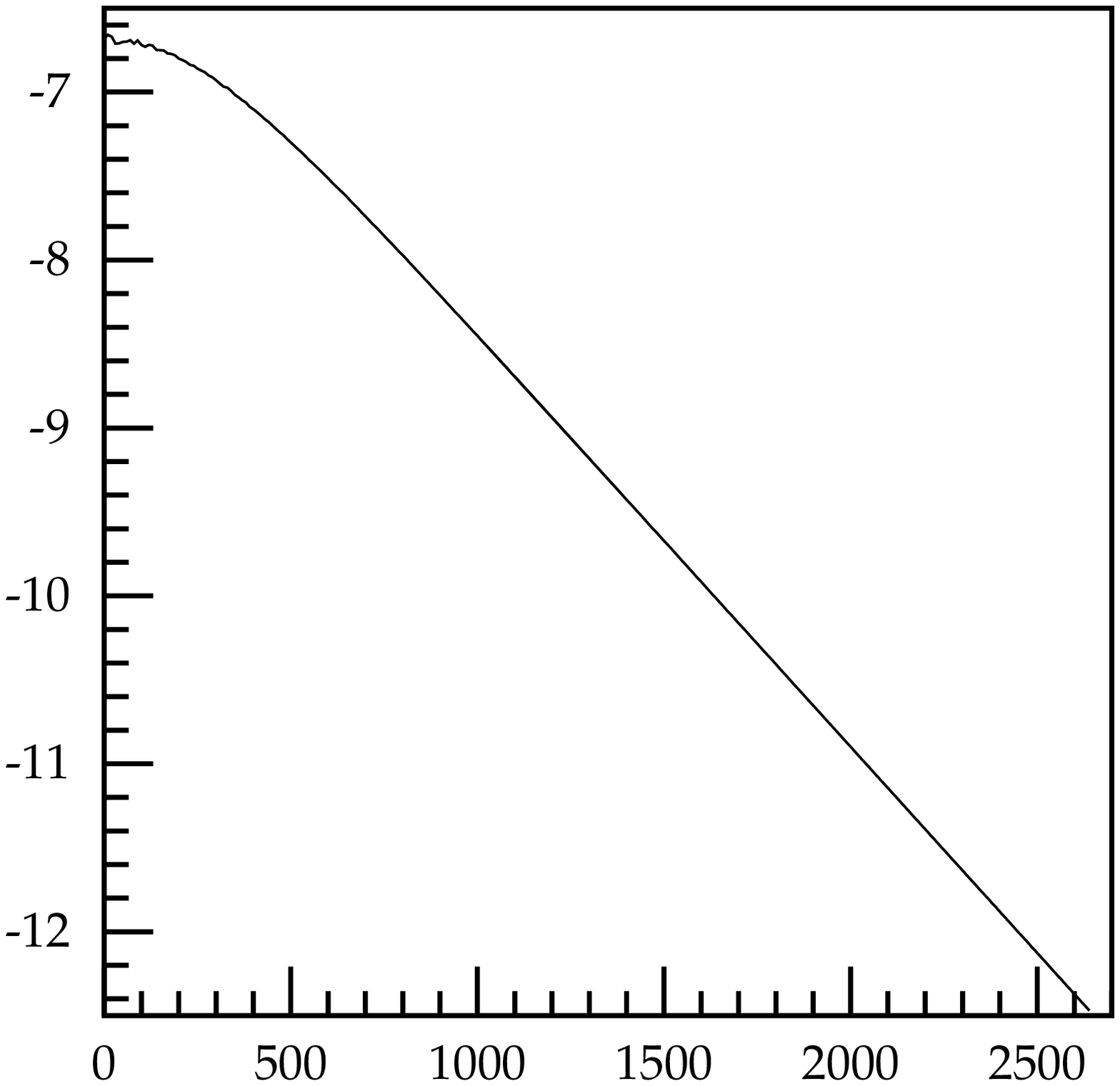}}
 \put(4,0){a}
\put(12,0){b}
\end{picture}
\caption{\label{traj_L10D} Trajectories of two kinks started with $d=6.5$ and at rest ({\it ie} with $v=0$) in two models of even $n$: a) $n=2$, b) $n=4$.}
\end{figure}

Of course, in the Sine Gordon model one has analytical expressions for moving kinks;
and it is not clear who to relate the numerical results to the analytical ones.

We plan to look at this in more detail in our further work.

\section{Acknowledgement}
So part of this work was performed when one of us (WJZ) was visiting the Eurasian National University. Astana, Kazakhstan. He wants to thank the Eurasian National University for its hospitality.

{}

\end{document}